\documentclass[prb,aps,amsmath,amssymb,showpacs,twocolumn]{revtex4-2}
\usepackage{amsmath,amssymb,amsfonts,graphicx,float,dcolumn,bm,color,colortbl,multirow}
\usepackage[titletoc,title]{appendix}
\usepackage[dotinlabels]{titletoc}
\usepackage{cancel}
\usepackage{braket}
\usepackage[caption=false]{subfig}
\usepackage{mathrsfs}
\usepackage{dsfont}
\usepackage[mathscr]{euscript}
\usepackage[hidelinks]{hyperref}
\usepackage{comment}
\usepackage[dvipsnames]{xcolor}

\begin{document}
\title{Thermal Drude weight in an integrable chiral clock model}
\author{Sandipan Manna}\author{G J Sreejith}\affiliation{IISER Pune, Dr Homi Bhabha Rd, Pune 411008, India}\date{\today}
\begin{abstract}
We calculate the finite temperature thermal conductivity of a time-reversal invariant chiral $\mathbb{Z}_3$ clock model along an integrable line in the parameter space using tDMRG.
The thermal current itself is not a conserved charge, unlike in the XXZ model, but has a finite overlap with a local conserved charge $Q^{(2)}$ obtained from the transfer matrix.
We find that the Drude weight is finite at non-zero temperature, and the Mazur bound from $Q^{(2)}$ saturates the Drude weight, allowing us to obtain an asymptotic expression for the Drude weight at high temperatures.
The numerical estimates are validated using a sum rule for thermal conductivity.
On the computational side, we also explore the effectiveness of the ancilla disentangler in the integrable and non-integrable regimes of the model. We find that the disentangler helps in localizing the entanglement growth around the quench location, but the improvement is lesser in the non-integrable regime and at low temperatures.
\end{abstract}
\maketitle
\section{Introduction\label{sec:intro}}

One-dimensional quantum systems host many experimentally realizable and theoretically tractable models that exhibit a rich set of quantum phenomena, including phase transitions, quantum transport, Luttinger liquids, etc.
Transport setups that measure energy, spin or charge currents generated in response to weak external field gradients are widely used in characterization of such systems.
Related questions of dynamics in weakly non-equilibrium situations have garnered considerable theoretical interest \cite{bertini2021finite} and inspired the development of efficient algorithms \cite{Prosen2009, PhysRevA.92.052116, karrasch2013reducing, paeckel2019time} for their study. The spin-1/2 XXZ chain is an analytically tractable model that can be realized in certain one-dimensional quantum magnets \cite{vasiliev2018milestones,BELIK2004883}.  Attempts to understand their measured transport properties led to many early works on isolated XXZ chains and closely related bosonic and fermionic models. These studies have revealed a connection between integrability and anomalous transport applicable not just in magnets but also in experimentally realizable bosonic and fermionic systems \cite{kinoshita2006quantum,bloch2012quantum,hofstetter2018quantum,bernien2017probing}.

Based on the seminal works by Mazur and Suzuki which relate the time-averaged correlations with overlaps of operators and conserved charges \cite{mazur1969non,suzuki1971ergodicity}, it was argued \cite{zotos1997transport} that the conductivity of integrable systems such as the XXZ chain should have an anomalous zero frequency singularity (Drude weight).
XXZ model is associated with two physically accessible conserved charges - namely energy and total spin quantum number along the $z$ direction. The possibility of a finite Drude weight for thermal and spin conductivity, as well as broader features of associated transport, have been investigated in the XXZ model and its variants (including the cases with impurity \cite{brenes2018high,brenes2020eigenstate} or noise \cite{de2022subdiffusive}) over the past several decades \cite{karrasch2017proposal, prelovvsek2021coexistence,karrasch2012finite,karrasch2015spin,langer2011real,ljubotina2017spin}.

In the gapless phase of the XXZ model, corresponding to the easy plane anisotropic case, early studies using Bethe ansatz suggested a finite zero temperature spin Drude weight which monotonically decreases with temperature \cite{zotos1999finite}. 
The spin Drude weight decreases towards zero at all temperatures as the isotropic Heisenberg point is approached \cite{urichuk2019spin}. 
Numerical studies using exact diagonalization and a finite temperature generalization of the Kohn's formula \cite{kohn1964theory,castella1995integrability,PhysRevB.58.R2921} hints at vanishing spin Drude weight at finite temperatures in gapped phase when the system has net zero magnetization \cite{peres1999curvature,herbrych2011finite,rabson2004crossover,prelovvsek2004anomalous}. Results using generalized hydrodynamics and thermodynamic Bethe ansatz approach also indicate zero spin Drude weight in the gapped phase and finite spin-Drude weight in the gapless phase \cite{urichuk2019spin}. 
Broadly the results also agree with tDMRG calculations which can access large finite size system sizes \cite{karrasch2012finite}.
These results pertain to the zero magnetization ensembles; the spin Drude weight is finite in ensembles with finite magnetization even in the gapped phase \cite{herbrych2011finite}.
Though the spin Drude weight has been found to be finite in the gapless phase, contrary to the expectation, the spin current operator is orthogonal to all local conserved charges obtained from the transfer matrix \cite{grabowski1995structure,zotos1997transport}. 
The finite Drude weight has been understood in terms of the Mazur bound related to the overlap of the spin current with certain additional quasi-local conserved charges \cite{prosen2013families,sirker2011conservation}.

The thermal current operator in the XXZ model is a conserved charge \cite{zotos1997transport}, and consequently the dynamical thermal conductivity has no finite frequency contribution. 
For the gapped phase, numerics and Bethe ansatz \cite{sakai2003non,zotos1997transport,heidrich2005thermal} approaches reveal a power-law decay with temperature of thermal Drude weight at high temperatures and an exponential decay with $1/T$ at low temperatures, separated by a finite maximum. The trend is similar for the gapless regime, but the low temperature behavior\cite{klumper2002thermal} is linear in $T$.

Besides the calculation of the Drude weight, other related aspects of ballistic transport have also been explored. These include studies on Fourier's law violations in non equilibrium steady state currents generated in response to a potential gradient \cite{prosen2009matrix,ajisaka2012nonequlibrium,vznidarivc2011spin,mendoza2013heat,mendoza2019asymmetry,nishad2022energy,brenes2018high,brenes2020eigenstate}, studies of currents in ray-dependent steady states after quench from a bipartitioned system  \cite{PhysRevB.98.075421,karrasch2013nonequilibrium,nozawa2020generalized,kormos2017inhomogeneous,zotos2017tba,bertini2016determination,doyon2017drude}, and studies on the nature of spread of initial spin or energy packets \cite{langer2011real,PhysRevB.79.214409,karrasch2014real,jesenko2011finite}. Breaking the integrability in such models, for instance, by next nearest neighbor interactions or noise, results in vanishing of the Drude weight and transfer of the conductivity weight to finite frequencies \cite{zotos1996evidence,PhysRevB.102.184304,de2022subdiffusive,heidrich2004transport,rabson2004crossover}. 

While the XXZ model and equivalent fermionic and bosonic models have been explored in great detail, not much has been systematically explored in other models \cite{metavitsiadis2017thermal}. In this work, we investigate the $\mathbb{Z}_3$ chiral clock model with a spatial chirality \cite{howes1983quantum,fendley2012parafermionic,ostlund1981incommensurate,huse1981simple}. The model hosts a rich phase diagram \cite{zhuang2015phase} and exhibits dynamics \cite {nishad2021postquench,jermyn2014stability,PhysRevB.98.075421} of interest in Rydberg atom systems \cite{fendley2004competing}. More importantly, for our current interest, in parameter regime the model is integrable \cite{AuYang1987CommutingTM}. 

We numerically calculate the dynamical thermal conductivity in the integrable regime using techniques developed and applied successfully in the XXZ model and characterize the Drude weight as a function of the temperature and model parameters. 
The Drude weight is found to be finite at all temperatures and the temperature dependence qualitatively matches that of the thermal Drude weight in the gapped phase of the XXZ model.
We then borrow results on the transfer matrix of the classical chiral clock model \cite{AuYang1987CommutingTM} to construct local conserved charges for the quantum model. The Mazur bound from the simplest among these charges $Q^{(2)}$ is found to saturate the numerically calculated Drude weight, which indicates that the current has a finite overlap only with $Q^{(2)}$. This allows us to obtain an asymptotically exact expression for Drude weight in the large temperature limit. The thermal current is not a conserved quantity, unlike the XXZ model and has a component that is not conserved and therefore orthogonal to all other conserved charges, we expect this to result in coexistence of diffusive and ballistic components in the dynamics of energy packets in this model. 

In the past, several numerical techniques have been utilised to address the questions on transport phenomena in interacting quantum systems. Exact diagonalization techniques can estimate correlator of local observables with high accuracy but suffers from severe finite-size effect making the results unreliable in thermodynamic limit \cite{long2003finite,prelovvsek2004anomalous,steinigeweg2014spin,heidrich2002thermal,heidrich2003zero,herbrych2011finite,rabson2004crossover}. Tensor network based tDMRG method have also been used to investigate large system sizes and therefore, practically approaching thermodynamic limit with tunable error and computation cost \cite{paeckel2019time,schollwock2011density}. A modified tDMRG approach based on the ancilla purification is found to be able to significantly reduce  the computational cost of the simulation \cite{karrasch2014transport,karrasch2012finite}. We use the latter in our calculation of the relevant correlators.

This paper is organized as follows. In Sec. \ref{THE QUANTUM Z3 CHIRAL CLOCK MODEL}, we introduce the model. Next, in Sec. \ref{Energy density section}, we present the expressions for the energy density and energy current; and introduce the relation between the conductivity and the current-current correlator which we use in the numerical calculations.
We also discuss constraints satisfied by the thermal conductivity namely the sum rules and Kramers-Kronig like relation which we use in subsequent sections to validate our numerical results. 
We discuss the Mazur bound and the conserved charges in Sec. \ref{Mazur bound}. Derivation of the conserved charges from the transfer matrix is presented in the Appendix \ref{app:TransferMatrix}.
In Sec. \ref{sec:Numerical Method}, we outline the tDMRG method that was used to obtain the numerical results. Finally, in Sec. \ref{sec: Numerical results}, we present our key results including the Drude weights as a function of temperature and parameters and comparison with estimates from Mazur bound. This section also discusses the frequency content of the regular part of the dynamical thermal conductivity. In Sec. \ref{Efficiency analysis of the modified tDMRG}, we present our empirical observations regarding the efficiency of the ancilla disentangler used in tDMRG calculations and conclude in Sec \ref{Summary and Outlook}.

\section{THE QUANTUM $\mathbb{Z}_3$ CHIRAL CLOCK MODEL}
\label{THE QUANTUM Z3 CHIRAL CLOCK MODEL}
In this section, we introduce the $\mathbb{Z}_3$ symmetric quantum chiral clock model in one spatial dimension \cite{howes1983quantum,ostlund1981incommensurate,huse1981simple}. We assume open boundary conditions throughout unless mentioned otherwise. The model can be motivated as a generalization of the $\mathbb{Z}_2$ symmetric transverse-field Ising model \cite{fendley2012parafermionic}.
The Hamiltonian is given by
\begin{equation}
	\label{Hamiltonian}
	H=- \sum_{j=1}^{L}[\bar{\alpha}_1  \tau_j+\bar{\alpha}_2  \tau_j^{\dag}+\alpha_1  \sigma_j \sigma_{j+1}^{\dag}+\alpha_2  \sigma_j^{\dag} \sigma_{j+1}]
\end{equation}
where the operators ${\tau}$ and ${\sigma}$ satisfy the following algebra,
\begin{equation*}
	\tau^{3}=\sigma^{3}= \mathbb{I}; \quad\tau^{\dagger}=\tau^{-1};\quad\sigma^{\dagger}=\sigma^{-1};\quad\sigma\tau=\omega\tau\,\sigma .
\end{equation*}	
and $\omega\equiv\exp\left(\frac{2\pi \imath}{3}\right)$. These can be explicitly represented by the following matrices.

\begin{equation} 
	\sigma=\left(\begin{array}{ccc}
		1 & 0 & 0 \\
		0 & \omega & 0 \\
		0 & 0 & \omega^{2}
	\end{array}\right), \quad \tau=\left(\begin{array}{lll}
		0 & 0 & 1 \\
		1 & 0 & 0 \\
		0 & 1 & 0
	\end{array}\right).
\end{equation}
The algebra of the operators $\sigma$ and $\tau$ generalizes that of the Pauli matrices $S_z$ and $S_x$ which appear in the Ising model. 

The Hamiltonian is parameterized by complex scalar parameters $\alpha_1$ and $\bar{\alpha}_1$, whose  complex conjugates are denoted by $\alpha_2$ and $\bar{\alpha}_2$ respectively. We emphasize that the bar does not indicate complex conjugates and the convention follows the one that has been used in literature \cite{AuYang1987CommutingTM}.

Alternatively, we can use the following parametrization. 
\begin{align}
	\label {alpha definition}
	\alpha_1&=J e^{\imath \theta}, \ \alpha_2=J e^{-\imath\theta} \nonumber\\  \quad
	\bar{\alpha}_1&=fe^{\imath\phi} , \ \bar{\alpha}_2 =fe^{-\imath\phi} 
\end{align}
where $f\geq 0$ is  local Zeeman field and $ J\geq 0 $ is nearest neighbor coupling. $\theta$ is a scalar parameter that governs the chirality. $\phi$ breaks the time-reversal symmetry.

Depending on the parameters $\theta$, $\phi$, the model can have discrete symmetries e.g. charge conjugation ($\sigma \leftrightarrow \sigma^{\dag}, \tau \leftrightarrow \tau^{\dag}$), spatial parity($j \leftrightarrow L-j+1$) and time-reversal ($H \leftrightarrow H^{*}$) symmetry. 
In this work, we focus on  $\theta \neq 0$ and $\phi = 0$ regime (i.e. $ \bar{\alpha}_1= \bar{\alpha}_2 =f\in \mathbb{R}$) where the model exhibits time-reversal symmetry. The model has a spatial chirality in this regime as the energy of left and right domain walls are not equal. The $\mathbb{Z}_3$ chiral clock model has a phase diagram with a disordered ($f\gg J$) phase and a $\mathbb{Z}_3$ symmetry broken ordered phase ($J\gg f$); the latter contains a topological regime \cite{zhuang2015phase,everts1989transfer,PhysRevB.88.085115}. 
The phases are separated by a surface of second order quantum phase transition points that passes through the critical three state Potts model at $J=f$, $\theta=\phi=0$ \cite{SachdevNumericZ3}. 

The model is integrable along the line \cite{AuYang1987CommutingTM}
\begin{align}
\label{integrability Jf}
f\cos(3\phi)=J\cos(3\theta)
\end{align}
and superintegrable when $\theta = \phi = \frac{\pi}{6}$ \cite{albertini1989commensurate,mccoy1990excitation}. This work aims to understand aspects of the model along the integrable line. For simplicity we consider only the time reversal invariant models and study the properties as $\theta$ varies keeping $\phi$ to be $0$, leaving the calculations in the finite $\phi$ regime for a future work.

\section{Energy density,Thermal current operator and Drude weight}
\label{Energy density section}
A key signature of an integrable model is the non-zero Drude weight which refers to a zero frequency delta function peak in the conductivity. In this work we explicitly calculate the Drude weight in thermal conductivity as a function of temperature and Hamiltonian parameters.
In the following subsections, we present the expressions for the energy density, thermal currents, conductivity and its Drude weight.

\subsection{Local energy density and thermal current operator\label{sec:energyDensity}}
In order to calculate the Drude peak (and dynamical thermal conductivity in general), we study the two-point correlator for the thermal current whose definition depends on the choice of the form of the local energy density operator. 
We choose the local energy density as a 3-site operator,
\begin{equation}
	\label{potts energy density}
	H_{i}=-\frac{\alpha_1}{2}(\sigma_{i-1}\sigma_{i}^{\dagger}+\sigma_{i}\sigma_{i+1}^{\dagger})-\bar{\alpha}_1\tau_{i}+h.c
\end{equation}
The thermal current operator, $I_i$ acting on the bond between site $i$ and site $i+1$ can be inferred from continuity equation as $	\dot{H_i}=\imath [H,H_i]=-\{I_{i+1}-{I}_{i}\}$. The local current operator $I_i$ has the form
\begin{equation}
	\label{potts current}
	I_{i}=\imath \frac{\bar{\alpha}{\alpha_1}}2\bigl(I_{i}^{(1)}+ I_{i}^{(2)}\bigr)+h.c
\end{equation}
where
\begin{equation}
	\label{current_potts_sub def}
	\begin{aligned}
		I_{i}^{(1)}=(\omega^{2}-1)\sigma_{i}(\tau_{i}+\tau_{i+1}){\sigma_{i+1}^{\dag}}\\
		I_{i}^{(2)}=({\omega}-1)\sigma_{i}({\tau_{i}^{\dag}}+{\tau_{i+1}^{\dag}}){\sigma_{i+1}^{\dag}}.
	\end{aligned}
\end{equation}
The total current operator is defined as $I = \sum_{i=1}^{L}I_i$. Symmetry considerations imply that the current operator has zero expectation value in any thermal state \cite{nishad2022energy}. 

\subsection{Thermal conductivity and sum rule}
Fourier transform of thermal conductivity in the linear response regime is given by \cite{luttinger1964theory,kubo1957statistical}
\begin{equation}
\label{conductivity linear response}
\kappa(k,\Omega_{+})={\frac{1}{\hbar T k{L}}}\int_{-\infty}^{0}\,e^{-\imath\Omega_{+}t}\langle[{\hat{I}}(k,0),{\hat{H}}(-k,t)]\rangle\,{\rm d} t
\end{equation}
where operators with hat ($\hat{I}$ and $\hat{H}$) represent the corresponding spatial Fourier transformed form,  $\Omega_{+}=\Omega + \imath 0^+$, $T$ is temperature, $L$ is the number of sites. Integrating Eq. \ref{conductivity linear response} by parts and subsequently using continuity equation and the KMS condition \cite{haag1967equilibrium} on the resulting expression, the thermal conductivity in the long wavelength limit ($k\to 0$) reduces to 
\begin{equation}
	\label{conductivity current current correlation}
	 \kappa(\Omega) =\frac{1-e^{-\beta \Omega}}{T  \Omega}\int_{0}^{\infty}e^{\imath\Omega t} \lim_{{L}\to \infty} \frac{\langle I(t)I(0) \rangle}{L} {\,\rm d}t
\end{equation}
which can be used to numerically evaluate the thermal conductivity (hereafter we use thermal conductivity to refer to its long wavelength limit only). The current-current correlator is directly accessible in the numerical time evolution which we implement using finite temperature tDMRG methods \cite{schollwock2011density,karrasch2012finite,karrasch2013reducing,karrasch2014real,karrasch2013nonequilibrium,barthel2013precise}. 

As shown in Ref. \onlinecite{{shastry2006sum}} thermal conductivity satisfies the following sum rule:
\begin{equation}
	\label{eq:sumRule}
	\int_{0}^{\infty}\mathrm{R e}\;\kappa(\Omega){\,\rm d}\Omega=\lim_{L\to\infty} \frac{\pi}{2\hbar TL}\langle\Theta\rangle
\end{equation}
which we can user to validate the results of the numerically estimated conductivity. We set $\hbar=1$ throughout this work. Here $\Theta$ is called the thermal operator for which an exact expression is given by,
\begin{equation}
	\label{theta def}
	\Theta=-\operatorname*{lim}_{k\rightarrow0}\frac{{\rm d}\hphantom{k}}{{\rm d} k}[{\hat{I}}(k),{\hat{H}}(-k)].
\end{equation}
The expectation value in Eq. \ref{eq:sumRule} is calculated in the thermal ensemble. For the specific choice of current and energy operator, $\Theta$ can be expressed in terms of local real space operators as follows. 
\begin{align}
	\label{thetaxx numerical definition}
	\Theta=
				\frac{\imath}{2}\sum_i [I_i,-3H_{i-1}-H_{i}+H_{i+1}+3H_{i+2}]
\end{align}

\subsection{Drude weight \label{drude weight}}
The extensive number of conservation laws in the integrable model can result in a non-zero value for the current-current correlator at asymptotically large time, resulting in a divergent zero frequency contribution to the thermal conductivity. The conductivity in the Fourier space has the following decomposition:
\begin{equation}
\label{conductivity Drude}
\kappa(\Omega) = 2\pi D_{\rm th}\delta(\Omega) +\kappa_{\rm reg}(\Omega)
\end{equation}
where,
\begin{align}
	\label{Drude wt formula}
	D_{\rm th}(T)&=\lim_{t \to \infty} \lim_{{L}\to \infty} \frac{ \langle I(t) I(0) \rangle}{2LT^2} \nonumber \\
	&=\lim_{t \to \infty} \lim_{{L}\to \infty} \frac{ \langle I(t) I_{L/2}(0) \rangle}{2T^2}
\end{align}
We have invoked spatial translation invariance in the second equality. This can be directly evaluated in tDMRG calculations.

As shown in Ref. \onlinecite{karrasch2014transport}, thermal conductivity can be calculated from the real or the imaginary part of the current-current correlator; and the difference between the two can be used as an estimate of the error in the calculated Drude weight. First, the real regular part of the dynamical thermal conductivity (Eq. \ref{conductivity Drude}) can be obtained directly from Eq. \ref{conductivity current current correlation}. Moreover, thermal conductivity $\mathrm{Re}\ \kappa_{\rm reg}(\Omega)$ can also be estimated using 
\begin{multline}
	\label{reg imaginary conductivity current current correlation}
	\mathrm{Re}\ \kappa_{\rm reg}(\Omega) =\\-\frac{2}{T  \Omega} \mathrm{Im} \int_{0}^{\infty}e^{\imath\Omega t}\, \mathrm{Im}\left[\lim_{{L}\to \infty}\frac{\langle I(t)I(0) \rangle_{\rm reg}}{L}\right] {\,\rm d}t
\end{multline}
In addition to the sum rule, discussed in the previous section, we use this to estimate error in the calculated conductivity and Drude weights (See Appendix \ref{app:consistency}).

\subsection{Mazur bound}
\label{Mazur bound}
The Drude weight is related to conserved charges of the integrable model through Mazur's inequality \cite{mazur1969non,suzuki1971ergodicity}. Assuming that the correlator has a steady asymptotic value, we can express this as the long time average of current-current correlator expectation value. For averages over a sufficiently large time window, the fluctuating contributions vanish from the spectral expansion, and we get \cite{sirker2020transport}
\begin{align}
	\label{integration correlation}
	&\lim_{\Lambda \rightarrow \infty} \frac{1}{\Lambda} \int_0^\Lambda dt \langle I(t)I(0) \rangle 
	&=\sum_{\substack{m,n\\E_n=E_m}}\frac{e^{-\beta E_n}}{Z} |\langle n |I|m \rangle|^2
\end{align}
where $Z={\rm Tr}[\rho]$, $\rho=e^{-\beta H}$ and $E_m$ denotes the energy of the $m^{\rm th}$ eigenstate, $|m\rangle$ and $\beta=\frac{1}{T}$ is the inverse temperature.

We denote the extensive set of Hermitian local conserved charges by $\{Q^{(j)} : [Q^{(j)},H]=0\}$. The conserved charges can always be made orthogonal (under the operator inner product $\langle A,B\rangle_{\rho}={\rm Tr}[\rho A^\dagger B]$) {\it i.e.} $\langle Q^{(j)},Q^{(k)} \rangle  \propto \delta_{jk}$.

Any operator that is diagonal in the energy basis must be a conserved charge for the Hamiltonian. This implies in particular that the diagonal part of current operator, $I$ is a constant of motion. Consequently, the total current operator, $I$ can be expanded as
\begin{align}
\label{conserved charge expansion}
I = \sum_{k=1}{} a_k Q^{(k)} +I'
\end{align}
where 
\begin{align*}
a_k=\frac{\langle Q^{(k)}, I \rangle_{\rho}}{\langle  Q^{(k)}, Q^{(k)} \rangle_{\rho}}
\end{align*}
and $I'$ is a purely off-diagonal matrix in any energy eigenbasis {\it i.e.} $\langle m | I' |n \rangle = 0 $ if $E_m = E_{n}$.
Using the expression for $I$ from Eq. \ref{conserved charge expansion} in Eq. \ref{integration correlation}, we obtain the following expression for the asymptotic value for the current-current correlator
\begin{align}
	\label{Mazur bound text}
	\lim_{\Lambda \rightarrow \infty} \frac{1}{\Lambda} \int_0^\Lambda  {\rm d}t \langle I(t)I(0) \rangle = \sum_{k} \frac{|  \langle Q^{(k)}, I \rangle_{\rho}|^2}{\langle Q^{(k)}, Q^{(k)} \rangle_{\rho}}
\end{align}
where sum is over conserved charges. Equation \ref{Mazur bound text} can be used to get the following expression for the Drude weight in terms of overlap of current operator and conserved charges:
\begin{align}
\label{Drude weight from conserved charge}
D_{\rm th}(T)&=\lim_{{L}\to \infty} \frac{1}{2LT^2} \sum_{k} \frac{|  \langle Q^{(k)}, I \rangle_{\rho}|^2}{\langle Q^{(k)}, Q^{(k)} \rangle_{\rho}} 
\end{align}
The right hand side is a sum of non-negative terms, and in the absence of a complete set of conserved charges the expression still provides a lower bound to the Drude weight.

There exists a one parameter family of $3^L$ dimensional transfer matrices $T(u)$ \cite{AuYang1987CommutingTM} parametrized by $u$ for the classical chiral Clock model in periodic boundary conditions. For parameters that satisify Eq. \ref{integrability Jf}, $T$ commutes with the quantum Hamiltonian as well as with each other ({\it i.e.} $[T(u),T(u')]=0$). 

An extensive number of local conserved charges for the quantum model can be obtained from a formal Taylor expansion of $\ln(T)$ in powers of $u$ (see for instance Ref \cite{Anjan1996}):
\begin{equation}
\ln T(u)=\sum_{j=1}^{\infty} Q^{(j)} u^j
\end{equation}
The simplest conserved charge $Q^{(1)}$ is the quantum Hamiltonian. The next term in the expansion namely $Q^{(2)}$ is a $3$-local conserved charge with the following form (See Appendix A for details)
\begin{multline}
	\label{Q2}
	Q^{(2)}=\sum_k  I_k+\imath\left(\frac{1}{2}+\omega \right)
	(\frac{ \alpha_1^2 \bar{\alpha}}{\alpha_2}-\bar{\alpha}^2)\tau_k+  \\
	 \imath \left(\frac{1}{2}+\omega^2 \right)
	(\frac{ \alpha_1^2 \bar{\alpha}}{\alpha_2}-\bar{\alpha}^2)\tau^{\dag}_k
 \end{multline}
Expression for $Q^{(3)}$ is presented in the Appendix \ref{app:TransferMatrix}.

The thermal current has a finite overlap with $Q^{(2)}$ and zero overlap with $Q^{(3)}$; this can be explicitly computed for the infinite temperature ensemble, and numerically seen to hold at all finite temperatures. This is similar to the case of the XXZ model \cite{zotos1997transport}.
Further, the numerical results in the subsequent sections show that the thermal current has a finite overlap only with $Q^{(2)}$ as the bound to the Drude weight calculated from only $Q^{(2)}$ using Eq. \ref{Drude weight from conserved charge} saturates the Drude weights calculated from asymptotic value of the finite temperature current-current correlator. However, unlike the XXZ model, the thermal current is not a conserved quantity in itself and $Q^{(2)}$ is not proportional the current. Only at critical point ($\alpha_1=\alpha_2=\bar{\alpha}$), conserved charge $Q^{(2)}$ coincide with the thermal current operator (Eq. \ref{Q2}).

\section{Numerical Method}
\label{sec:Numerical Method}
To calculate the dynamical current-current correlator in Eq.~\ref{Drude wt formula}, we employ a time-dependent DMRG method (tDMRG) implemented using matrix product states. 
We realize the finite temperature system as the subsystem of a bigger system containing the $L$ physical and $L$ ancilla degrees of freedom in a pure entangled state $\psi_{\beta}$ such that reduced density matrix of the physical degrees of freedom is $e^{-\beta H_{\rm{phy}}}/Z$ \cite{feiguin2005finite}. $H_{\rm{phy}}$ is the Hamiltonian in Eq. \ref{Hamiltonian} applied only on the subsystem of physical sites.
To prepare the state $\psi_\beta$ we start with a state where each physical site is maximally entangled with one neighbouring ancilla, representing the purification $\psi_{0}$ of an infinite temperature physical system.
The purification of the physical system at inverse temperature $\beta$ is obtained through an imaginary time evolution of the initial state: $\left|\psi_\beta\right \rangle \propto e^{-\frac{\beta}{2} H_{\rm{phy}}}\otimes \mathds{1}_{\rm{anc}}\left | \psi_0 \right \rangle $.
Both imaginary and subsequent real time evolutions are implemented through a fourth-order Trotter decomposition \cite{paeckel2019time}. The real-time evolution results in exponential growth of bond dimension in the MPS.

To slow down the bond dimension growth and access the largest possible time-scale, we exploit the fact that purification is not unique as outlined in Ref. \cite{karrasch2012finite}, due to the invariance of the reduced density matrix of the physical system under arbitrary unitary transformations of the ancilla subsystem. Entanglement growth can be slowed down by applying a suitable disentangler unitary on the ancilla.
For XXZ model, $U=\mathbb{I}_{\rm phys} \otimes e^{+\imath H_{\rm anc}t}$ where $H_{\rm anc}$ is the same as the Hamiltonian for the physical system ($H_{\rm{phy}}$) but, acting instead on the ancilla subsystem (See Ref \cite{barthel2013precise,karrasch2012finite,hauschild2018finding} for details).

In this model (Eq. \ref{Hamiltonian}) as well, we observe empirically that, within a restricted class of unitaries considered, the above unitary is optimal (See Appendix \ref{app:disentangler}).
This particular choice of $U_{\rm{anc}}$ makes time-evolution of thermal state $\left|\psi_\beta\right\rangle$ nearly trivial and for the quenched state (the thermal state perturbed by the operator {\it i.e.} $I_{L/2}\left|\psi_\beta\right\rangle$ in the present case),  
entanglement buildup gets confined in the vicinity of the site where the current operator is applied. We found that for our calculations it was sufficient to use a maximum bond dimension of $\chi$ of 900-1200 and a singular value cut-off ($\epsilon$) of $10^{-9}-10^{-12}$ for truncation to get results with sufficient convergence.
We repeated the calculation for multiple system sizes, cutoff bond dimensions ($\chi$) and truncation weight cutoffs ($\epsilon$) to confirm the convergence of our results (See Appendix \ref{app:convergence}). 

One further optimization often carried out in such numerical calculations of correlators is to  exploit the time translation invariance in correlator where we write $\langle A(t)B(0) \rangle$ as $\langle A(-t/2)B(t/2) \rangle$ \cite{kennes2016extending}. Unlike the case for the two-point correlator between a pair of {\emph{local}} observables, we have a task of evaluating a correlator of the form $\langle I(t)I_{\frac{L}{2}}(0) \rangle$ where one of the two operators is a {\emph{sum of local}} observables. 
The alternate strategy of calculating the required correlator as the sum of correlators of local objects $\langle I_i(t/2)I_{\frac{L}{2}}(-t/2) \rangle$ has a cost that is typically higher \cite{karrasch2013drude,kennes2016extending}.

All the numerical calculations in the following section were carried out using code built on ITensor \cite{itensor,itensor-r0.3}.
\section{Numerical results}
\label{sec: Numerical results}

\begin{figure}[h]
	\includegraphics[width=.8\columnwidth]{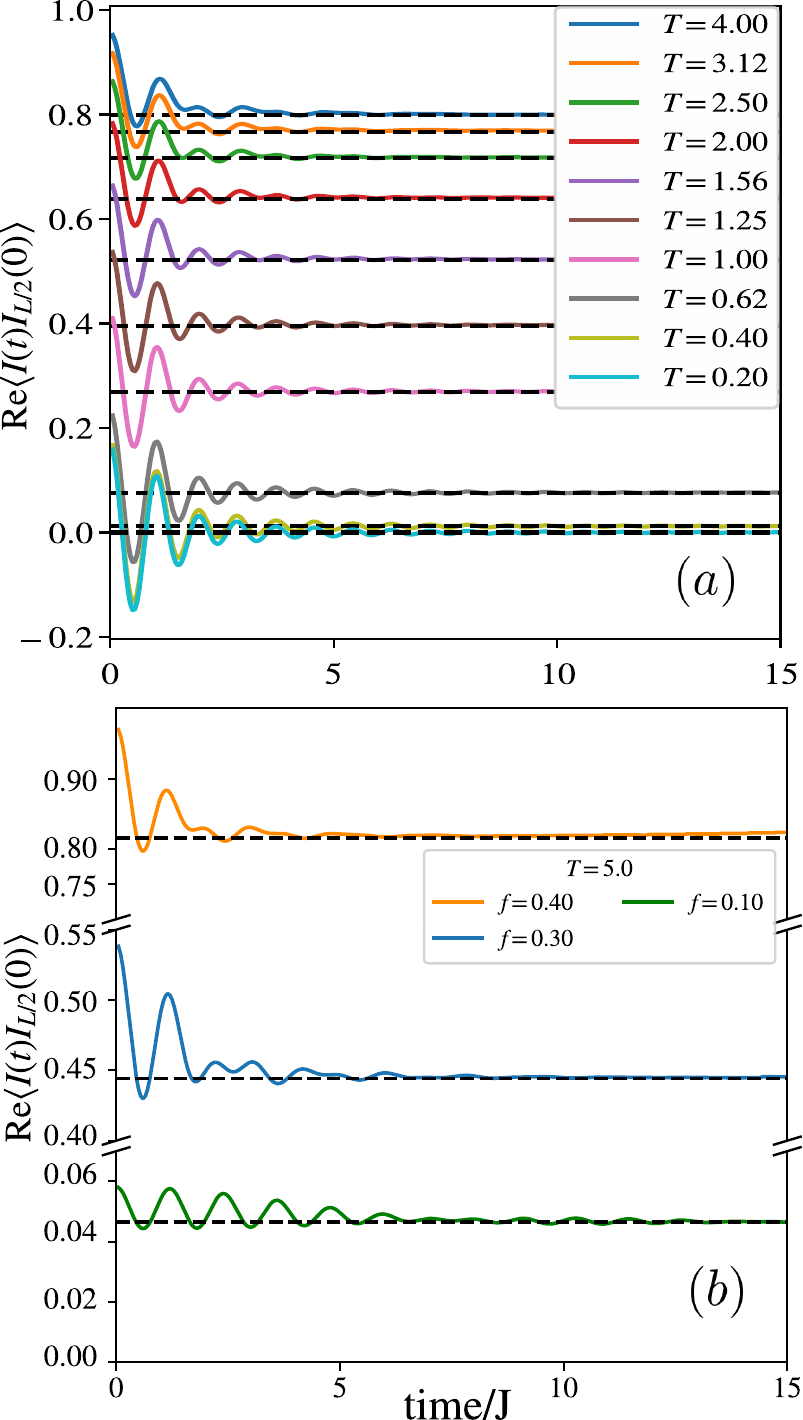}
	\caption{(a) Evolution of the current-current correlator. The dotted lines are lower bound on current-current correlator at different temperatures obtained from Mazur bound.  The maximum bond dimension for MPS is kept at 900-1200 with a cutoff for truncation $\sim 10^{-9}-10^{-10}$. $J=1, f=2/5, \theta=\frac{1}{3}\cos^{-1}(2/5)$. (b) Evolution of the current-current correlator for different $f$ along the integrable line at $T=5$. Cutoff for truncation, $\epsilon \sim 10^{-9}-10^{-12}$.  }
	\label{fig:JJ_beta_0_2}
\end{figure}

In this section, we present the results from the numerical calculations using methods discussed in the previous section. All the results presented in the main text are for a system of $L=60$ physical sites.
This section is organized as follows. The general features of the dynamical current-current correlations and thermal conductivity are presented in Sec. \ref{Evolution and convergence current-current correlation value} and verification of the sum rule is discussed in Sec. \ref{sec:sumrule}. 
The main results of the work namely the variation of Drude weight with temperature and parameters as well as a comparison with the Mazur bound are described in Sec. \ref{sec:DrudeWtResults}.
Lastly, we present an analysis of the efficiency of the disentangler unitary that we chose in  subsection \ref{Efficiency analysis of the modified tDMRG}. Except in the last subsection, we consider only the integrable line, along which the chirality $\theta$ is fixed by the integrability condition (Eq. \ref{integrability Jf}); as $f$ increases from $0$ to $1$, $\theta$ decreases from $\pi/6$ to $0$.

\begin{figure}[h]
	\includegraphics[width=.78\columnwidth]{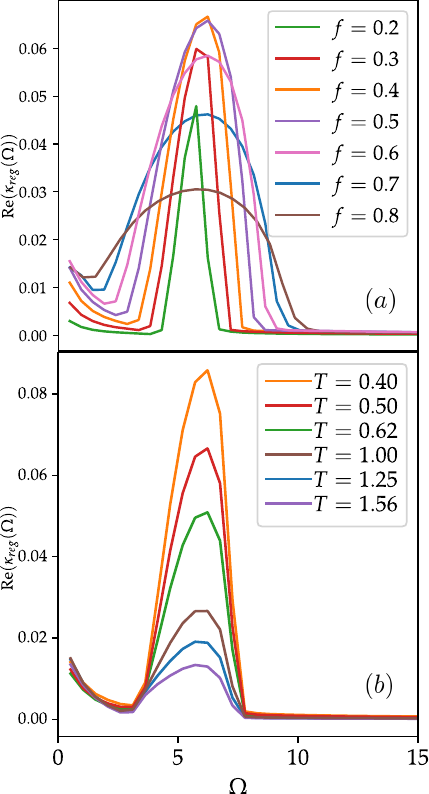}
	\caption{(a) Real part of Fourier transformed regular thermal conductivity $T=0.5$ as a function of $\Omega$. Different lines correspond to different values of Hamiltonian parameter $f$ along the integrable line.  (b) Similar to (a) but for different values of temperature ($T$). Results shown are for $J=1$, $f=2/5$, $\theta=\frac{1}{3}\cos^{-1}(2/5)$, $\phi=0$. }
        \label{fig:fourier_sigma_merged}

\end{figure}

\subsection{Current-current correlator and thermal conductivity\label{Evolution and convergence current-current correlation value}
}
Fig. \ref{fig:JJ_beta_0_2}(a) shows the time dependence of the real part of the current-current correlator for a range of temperatures at a fixed Hamiltonian parameter $f=2/5$. 
The dotted lines show the estimates from the Mazur bound (Eq. \ref{Mazur bound text}). The calculated current-current correlator is in excellent agreement with the Mazur bound. Generally the time taken to reach the asymptotic value is larger at lower temperatures.
Time dependence of the real part of the current-current correlator at a fixed temperature $T=5$ and different points along the integrable line is shown in Fig. \ref{fig:JJ_beta_0_2}(b). We find that the current-current correlator reaches its asymptotic value  faster at larger $f$.
We note that in the time series plot shown in Fig. \ref{fig:JJ_beta_0_2}(a,b) or the Fourier transforms presented  below, we have not used any extrapolation on the time series data \cite{barthel2009spectral} obtained from tDMRG. All the data shown are from the actual tDMRG evolution. We did not find any qualitative changes to the results upon using the extrapolated data.

Fourier transform of the conductivity can be calculated from the current-current correlator using Eq. \ref{conductivity current current correlation}. Fig \ref{fig:fourier_sigma_merged}(a) shows the variation of the regular part of the thermal conductivity as a function of $\Omega$ across a range of Hamiltonian parameter $f$. As $f$ increases from $0$ to $1$ (simultaneously $\theta$ varies from $\pi/6$ to $0$), the peak broadens but for $f$ close to  $1$ the peak rapidly decreases in height and vanishes at $f=1$. In particular the $\kappa_{\rm reg}$ is exactly $0$ for all frequencies at the critical Potts model point ($f=1,\theta=0$). 

 $\kappa(\Omega)$ has a delta function peak arising from the Drude weight, the remaining regular part $\kappa_{\rm reg}$ of the conductivity as a function of the frequency is shown in Fig. \ref{fig:fourier_sigma_merged}(b). $\kappa_{\rm reg}$ is near zero at the highest accessible frequency and is finite for $\Omega\to 0$. This finite DC value indicates a finite diffusive part to the energy transport in addition to the ballistic part. Due to errors arising from finite time data (See Appendix \ref{app:consistency}) we are not able to detect the temperature dependence of $\kappa_{\rm reg}(\Omega\to 0)$. At an intermediate range of frequencies ($3.0$ to $7.5$ in this case) there is a finite peak whose height decreases with increasing temperature. The frequency range is independent of the temperature.

Now we consider the location of the peak in relation to the spectral gap.
For the value of $f$ considered in Fig. \ref{fig:fourier_sigma_merged}, the gap above the ground state is $\Delta=1.2551$ (calculated using DMRG in a system of size 60). This should correspond to the lowest energy domain wall quasiparticle. Energy of the domain wall with the opposite chirality should be higher due to finite chirality $\theta$ \cite{NaveenEntanglement}.
Insertion of the current operator $I_{L/2}$ is expected to generate a pair of opposite chirality domain walls  ($\tau$ operators in Eq. \ref{current_potts_sub def} flips a single spin, generating a pair of domain walls around it) whose energy should be more than $\sim 2\Delta$ consistent with the low frequency end $\Omega_{\rm low}$ of the peak. An alternate possibility is that a chain of three domain walls $\left|...11...\omega\omega...\omega^2\omega^2...11...\right \rangle$ of same chirality costing an energy of $\sim 3\Delta$ determines the location of low frequency end. Within the resolution possible for $\Omega_{\rm low}$ the two scenarios cannot be distinguished.
As the critical point is approached, $f\to 1$, $\theta\to 0$, the spectral gap vanishes, which is consistent with vanishing $\Omega_{\rm low}$ in Fig. \ref{fig:fourier_sigma_merged}(a).

\begin{figure}[h!]
	\includegraphics[width=\columnwidth]{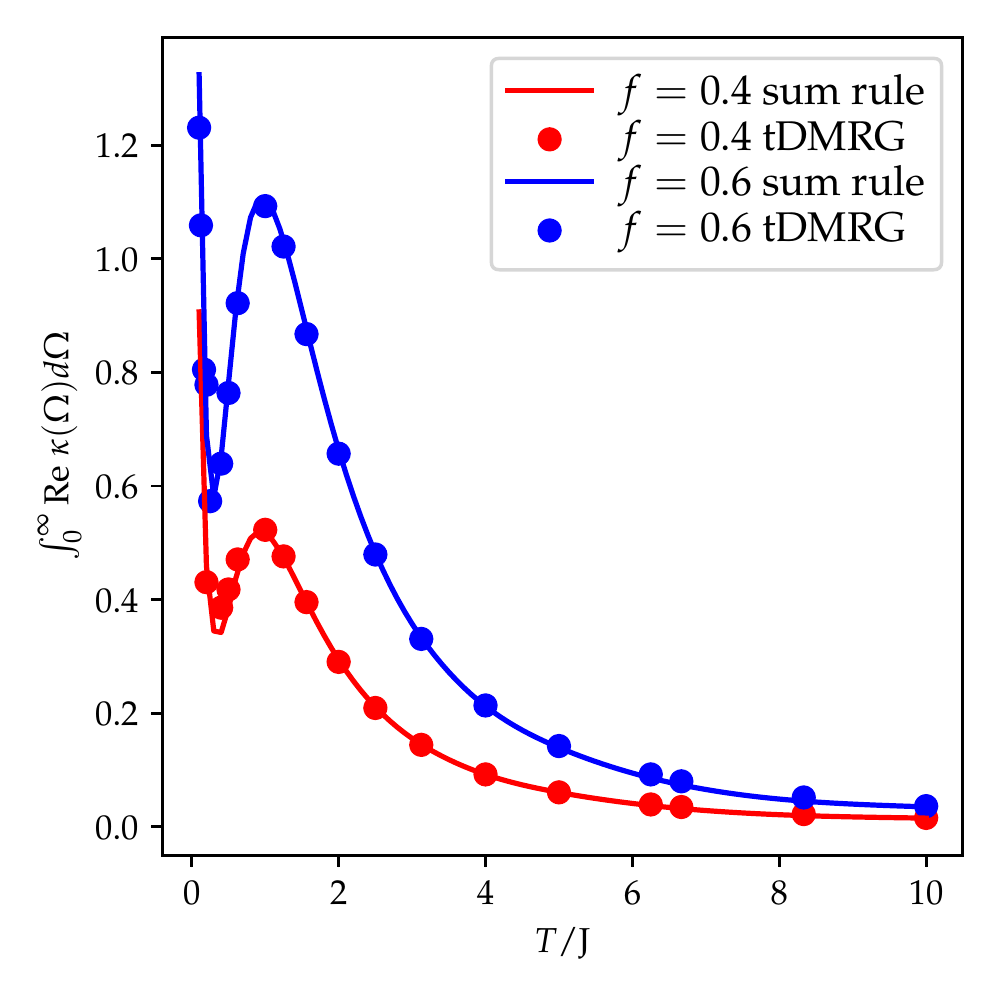}
	\caption{Verification of the sum rule by comparing RHS of Eq. \ref{eq:sumRule} (solid line) where the expectation value of $\Theta$ is calculated numerically using Eq.\ref{thetaxx numerical definition} . The dots indicate the integral of real part of Eq. \ref{conductivity current current correlation} obtained from real time correlatlor $\langle I(t)I_{\frac{L}{2}} \rangle$ evaluated using tDMRG. The different lines show results for different points on the integrable line.}
	\label{fig:temp_thetaxx}
\end{figure}

\subsection{Sum rule\label{sec:sumrule}}
We now verify the sum rule for thermal conductivity following Eq.\ref{eq:sumRule}. Fig. \ref{fig:temp_thetaxx} shows the integrated value of (numerically calculated) real part of thermal conductivity, $\rm{Re} \ \kappa(\Omega)$ (LHS of Eq. \ref{eq:sumRule} shown as dots in the figure) compared with the expected sum (RHS of Eq. \ref{eq:sumRule} shown as continuous line) as a function of temperature. We find good agreement between the two, implementing a partial check on the current-current correlators calculated numerically. 

\begin{figure}
\label{T_drude}
\includegraphics[width=\columnwidth]{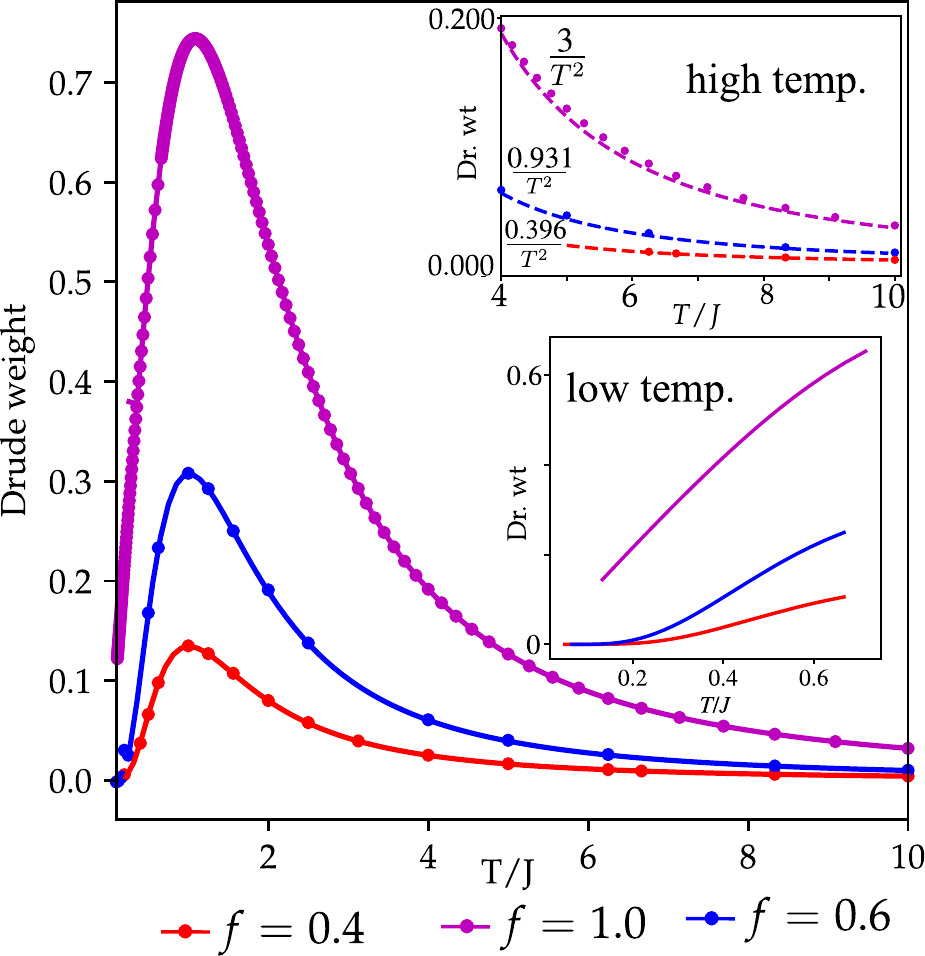}
\caption{Variation of the Drude weight with temperature for different values of $f$. The solid lines show the Mazur bound (Eq. \ref{Drude weight from conserved charge}). Dot markers show the Drude weight estimated from tDMRG data (Eq.\ref{Drude wt formula}). Temperature axis starts from $T=0.1$. (Inset) High temp: Asymptotic expression for Drude weight is represented by the dashed lines (Eq. \ref{D_th_inf}). The dots are from tDMRG estimates. Low temp: Drude weight from Mazur Bound in the low temperature regime. Behavior for the critical point is consistent with a linear scaling with $T$ while for the gapped phase the trend is consistent with $e^{-\frac{\delta}{T}}$.}  \label{fig:T_drude}
\end{figure}

The $T$ in denominator of Eq.~\ref{eq:sumRule} results in the divergence at low temperature in Fig. \ref{fig:temp_thetaxx}. 
The sharp increase in integrated thermal conductivity at low temperature arises from the intermediate frequency peak in Fig. \ref{fig:fourier_sigma_merged} that gets more significant at low temperatures. 
The finite temperature peak in the sum rule arises from the singular, zero frequency part {\it i.e.} Drude weight of the thermal conductivity. Fig. \ref{fig:temp_thetaxx} has a similar trend as in Fig.~\ref{fig:T_drude} at intermediate and high temperature.
\begin{figure}[H]
	\includegraphics[width=\columnwidth]{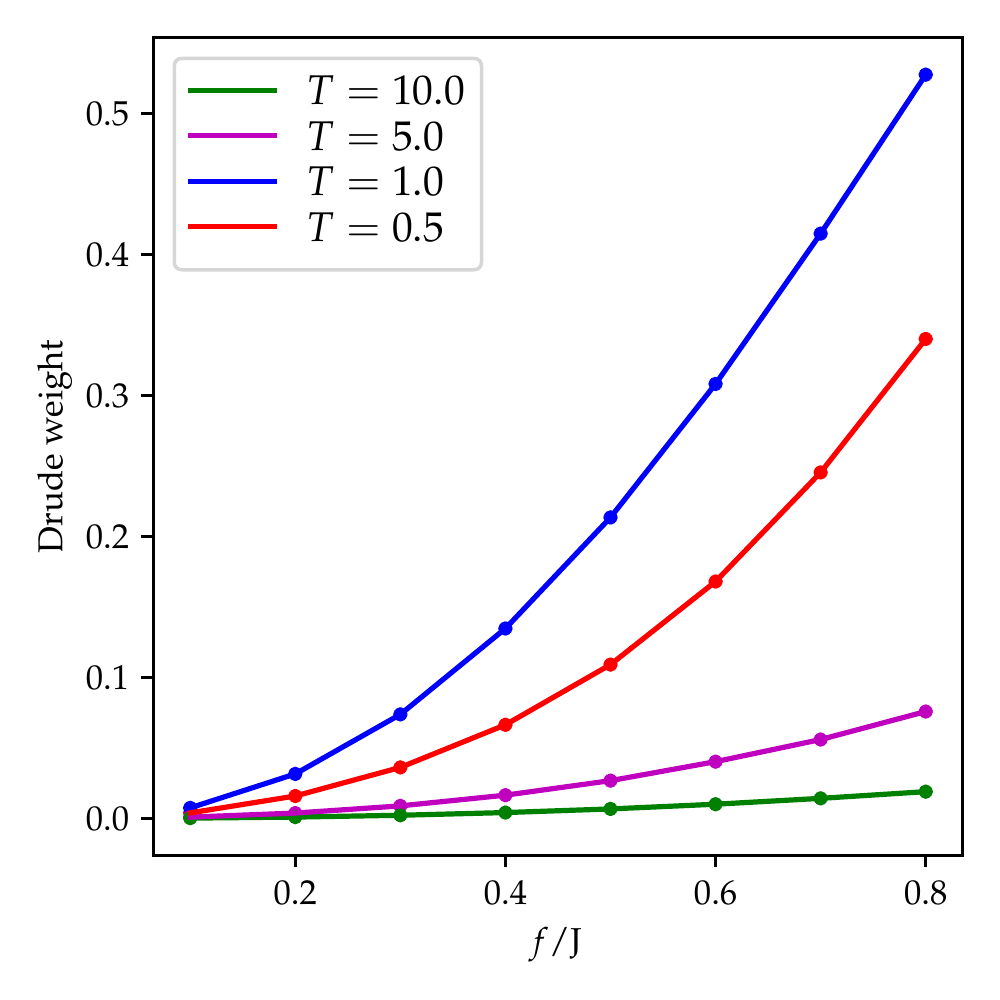}
	\caption{Drude weight calculated from tDMRG as a function of $f$ showing the monotonic  increase of Drude weight with increasing $f$. $\theta$ is chosen accoding to $f$ along integrable line. The solid line is a linear interpolation between data shown as a guide to the eye.\label{fig:f_drude}}
\end{figure}

\subsection{Variation of Drude weight with temperature and Hamiltonian parameters \label{sec:DrudeWtResults}}

In this section, we present the results for the numerically calculated Drude weights as a function of the temperature and comparison with the Mazur bound.
Fig \ref{fig:f_drude} shows the Drude weight computed from tDMRG as a function of the $f$ for different temperatures showing a monotonic (nearly quadratic at small $f$) increase with $f$. Drude weight changes non-monotonically with temperature as can be inferred from comparing the plots for $T=0.5,1,5.0$. 

Fig. \ref{fig:T_drude} shows the Drude weight numerically computed using tDMRG (dots in the figure) as a function of temperature for different values of $f$; and as mentioned before, the temperature dependence is non-monotonic. The solid line shows the Mazur bound calculated numerically using Eq. \ref{Drude weight from conserved charge} considering only the $Q^{(2)}$ charge (See discussion in Sec \ref{drude weight} and Appendix \ref{app:TransferMatrix}). The Mazur bound from $Q^{(2)}$ saturates the estimated Drude weight indicating that the thermal current has no overlap with any other conserved charge.

Drude weight is found to be zero at zero temperature and increases rapidly as temperature is increased till it reaches a finite temperature peak around $T=1$. The high temperature asymptotic behavior of the Mazur bound from charge $Q^{(2)}$ can be estimated to be
\begin{align}
	\label{D_th_inf}
	D_{\rm th}(T=\infty) = \frac{1}{2T^2}\frac{\langle I,Q^{(2)}_i \rangle_{\rho=\mathbb{I}}}{\langle Q^{(2)}_i, Q^{(2)}_i\rangle_{\rho=\mathbb{I}}} =\frac{1}{2T^2} \frac{24 f^2}{5-f^2}.
\end{align}
For $f\ll 1$, the above expression suggests a quadratic variation with $f$ consistent with Fig. \ref{fig:f_drude}. Figure \ref{fig:T_drude}(inset : high temp) shows the agreement (asymptotically) of Eq. \ref{D_th_inf} with data obtained from tDMRG at high temperature. 

In the low temperature (inset:low temp) regime, Drude weight decays linearly with $T$ for the critical system. The behaviour is qualitatively different in the gapped system ($f \neq 1$) where we observe a decay to 0 as $e^{-\frac{\delta}{T}}$. We can extract $\delta$ by fitting this to our data. For the two cases we considered here, $f=0.4$ and $0.6$, we find $\delta \sim 1.19$ and $0.98$  which are close to the corresponding spectral gap $1.25$ and $0.95$ above the ground state respectively.

\subsection{Efficiency analysis of the tDMRG with ancilla disentangler}
\label{Efficiency analysis of the modified tDMRG}

\begin{figure}[H]
\includegraphics[width=\columnwidth]{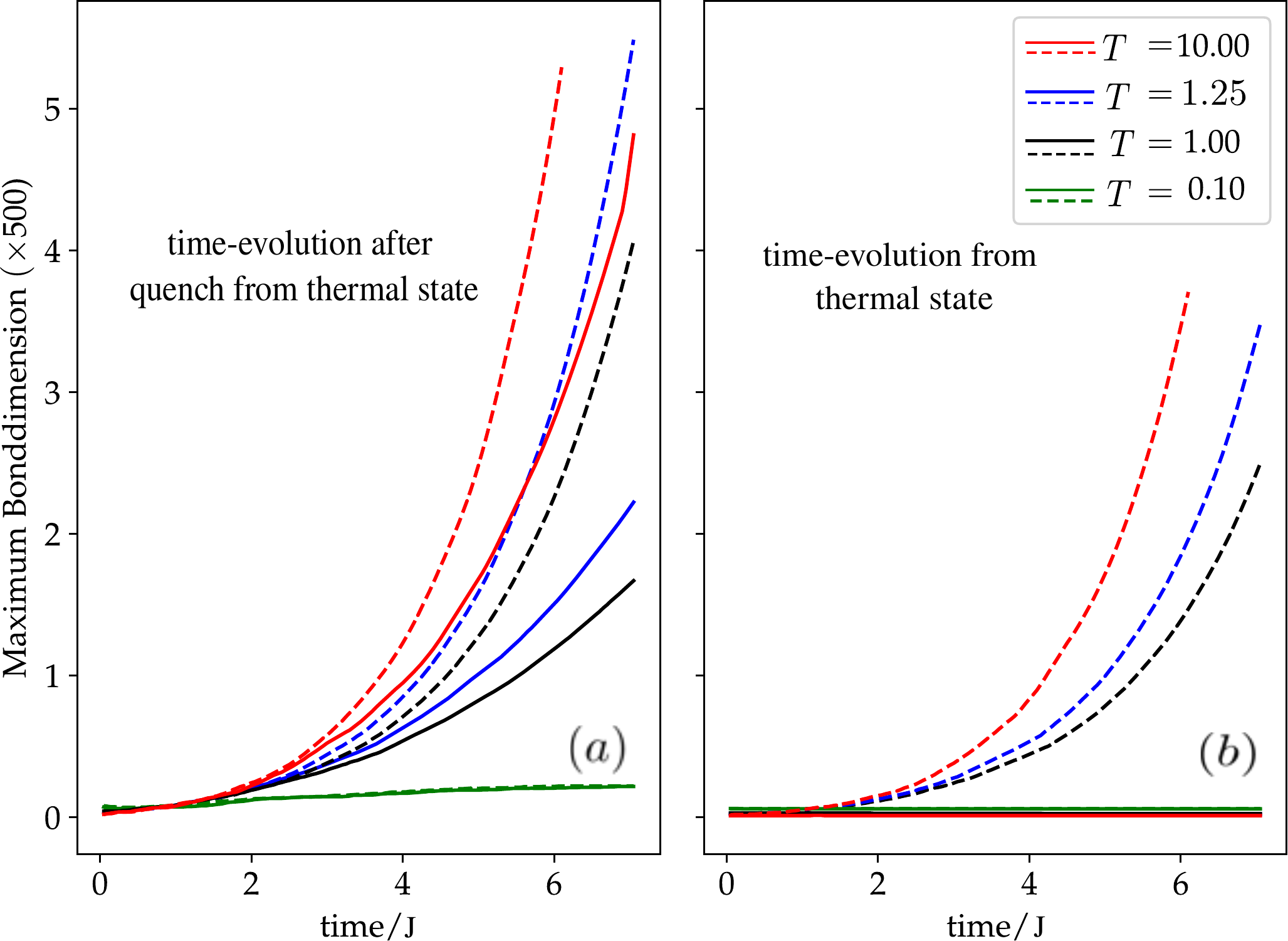}
\caption{(a)Growth of maximum bond dimension under real time evolution for the MPS quenched from thermal state at different values of $\beta$. The dotted line indicates tDMRG without application of disentangler and the solid line is for disentangled tDMRG. (b) Same as panel a but, for time-evolution of thermal MPS without any quench.The parameters for the Hamiltonian are $J=1$, $f=2/5$, $\theta=\frac{1}{3}\cos^{-1}(2/5)$, $\phi=0$ (integrable point). The truncation cutoff,$\epsilon = 10^{-9}$. \label{fig:bonddim_time}}
\end{figure}
The results presented in the previous sections were achieved using a disentangler unitary (See Sec. \ref{sec:Numerical Method}) on the ancilla subsystem in the tDMRG calculations. The disentangler slows down entanglement growth in general and makes accessing longer time scales possible. 
The effectiveness of this approach however varies with the parameter and temperature regimes; qualitatively similar to what was found in the case of XXZ model \cite{barthel2013precise}. In this section, we present our empirical observations along with some quantitative data on the efficacy of the disentangler approach across model parameters and temperature for the model studied here.

In Fig. \ref{fig:bonddim_time} (a) and (b) we show the variation of the maximum bond dimension after time-evolution of the quenched ($I_{\frac{L}{2}}|\psi_{\beta} \rangle$) and thermal ($|\psi_{\beta} \rangle$) states respectively. In each case we compare the bond-dimension growth with and without disentanglers applied to the ancilla.
From panel (a), we observe that the maximum bond-dimension of the quenched MPS grows much faster (up to 2.5 times) for tDMRG without disentangler (dotted line) in contrast to the tDMRG with disentangler (solid line).
At low temperatures, on the other hand, both methods show almost similar growth in maximum bond dimension with time. 
In Fig. \ref{fig:bonddim_time} (b), we see that the tDMRG with disentangler almost fully eliminates bond dimension growth of the thermal MPS under unitary evolution. In contrast, unitary evolution of thermal MPS ($|\psi_{\beta}\rangle$) implemented using tDMRG without disentangler shows rapid growth of maximum bond dimension with time.

\begin{figure}[h]
	\includegraphics[width=\columnwidth]{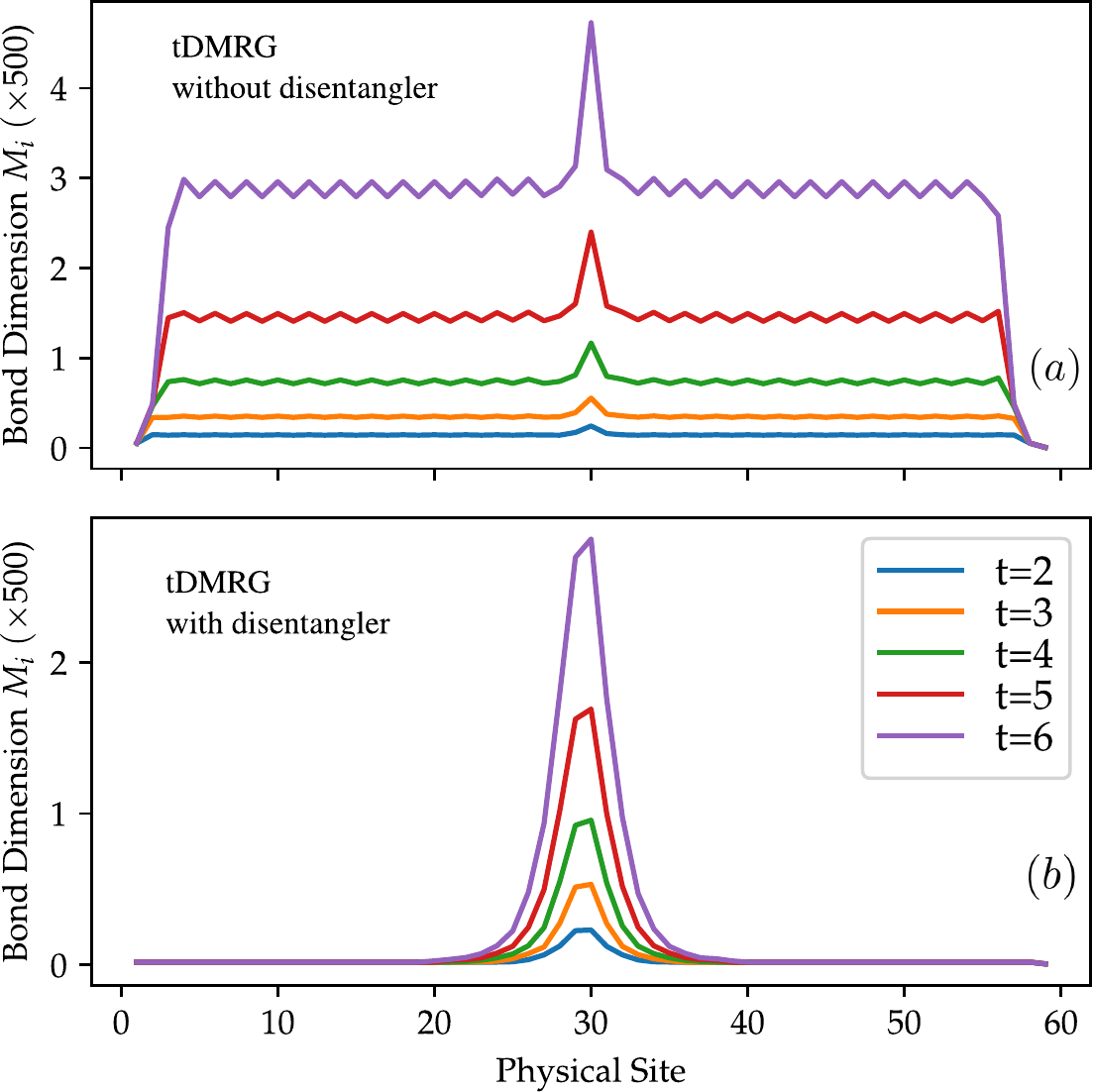}
	\caption{Bond dimension as a function of bond position for the state quenched from a high temperature ($T= 6.25J$) thermal state. Panel (a) data for tDMRG without disentangler and panel (b) shows data for tDMRG with disentangler. Different lines correspond to different times. The MPS bond dimension growth is localised about the quench site in case of tDMRG with disentangler. Results shown are for the integrable point $J=1,f=2/5$, $\theta=\frac{1}{3}\cos^{-1}(2/5)$, $\phi=0$. The truncation cutoff, $\epsilon = 10^{-9}$. }
	\label{fig:sitedata_beta_0_16}
\end{figure}

\begin{figure}[h]
	\includegraphics[width=\columnwidth]{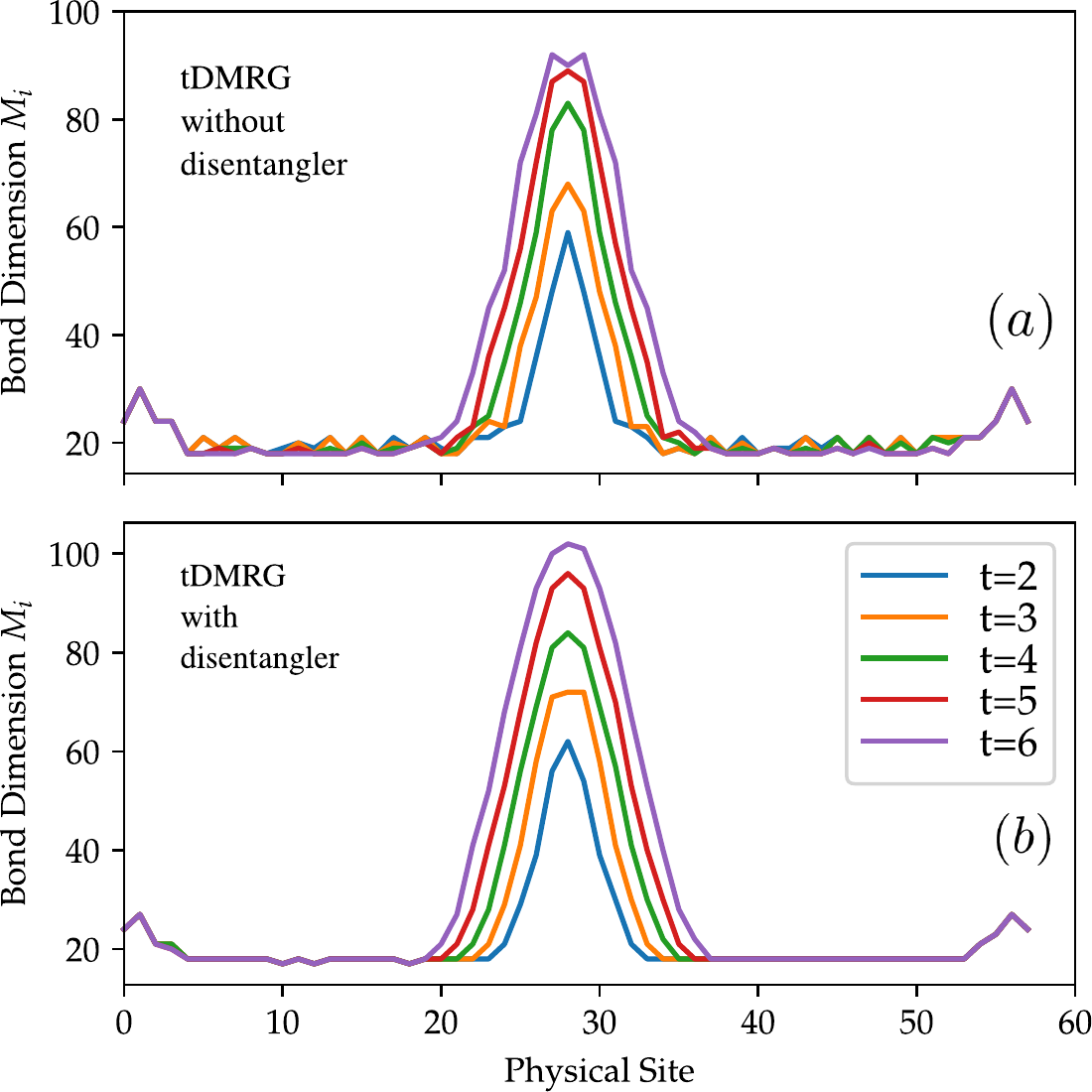}
	\caption{Similar to Fig. \ref{fig:sitedata_beta_0_16}, this plot shows the spatial distribution of bond dimension but at a low temperature point $T=0.1$ (in unit of $J$). The parameters for Hamiltonian are $J=1$ , $f=2/5$, $\theta=\frac{1}{3}\cos^{-1}(2/5)$, $\phi=0$. The truncation cutoff is $\epsilon = 10^{-9}$. }
	\label{fig:sitedata_beta_10_0}
\end{figure}

In Fig. \ref{fig:sitedata_beta_0_16}, we show the bond dimension of the time evolved states, as a function of the bond position contrasting the cases with and without disentangler. 
We notice a uniform growth of bond dimension throughout the system in the state that evolved without the ancilla disentangler in contrast to localized growth of bond dimensions near the quench site in case of tDMRG with disentangler. The spatial support of the operator is expected to spread linearly with time evolution; this increase is manifested in the excess growth of entanglement around the quench site (center in our case).
Without disentangler, however there is an additional uniform buildup of entanglement everywhere in the system.
In Fig. \ref{fig:sitedata_beta_10_0}, we show local bond dimension at low temperature. Here, in both cases (tDMRG with and without disentangler unitary), local bond dimension shows similar behaviour indicating no significant gain in efficiency from application of ancilla disentangler.

\begin{figure}[H]
	\includegraphics[width=\columnwidth]{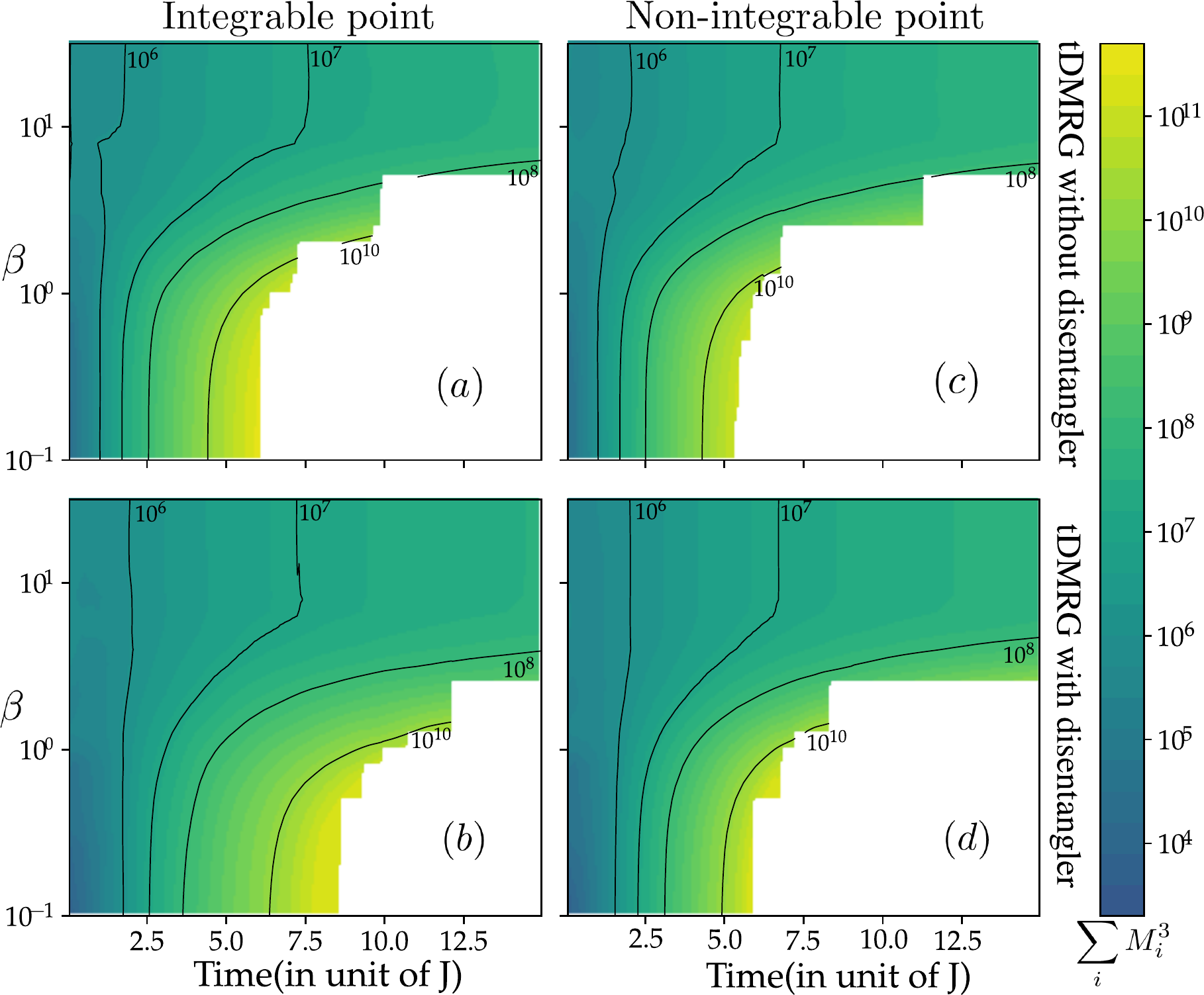}
	\caption{Computational complexity as a function of $\beta$ and time. (a) and (b) are for integrable model ($J=1$, $f=2/5$, $\theta=\frac{1}{3}\cos^{-1}(2/5)$, $\phi=0$) with and without disentangler unitary respectively. (c) and (d) are for non-integrable point $J=1$, $f=2/5$, $\theta=0.3$, $\phi=0$ with and without disentangler unitary respectively. Colors indicate a measure of the computational cost. Contour lines connect points of same computational cost. The truncation cutoff was maintained at $\epsilon = 10^{-9}$ throughout the calculations.}
	\label{fig:contour}
\end{figure}

We, so far, focused on the local bond dimension growth. However, time evolution calculations are dominated by matrix multiplications and SVD operations which have a computational complexity of $O(d^3)$ where $d$ is the dimension of the matrices involved. Denoting the local bond dimension at site $i$ of the MPS by $M_i$, the computational cost can be quantified by $\sum_i M_i^3$ \cite{barthel2013precise}. In Fig. \ref{fig:contour}, we provide a contour plot showing the variation of the computational cost $\sum_i M_i^3$ with inverse temperature $\beta$ and time, while keeping the truncation cutoff fixed at $10^{-9}$. 

We first compare Fig. \ref{fig:contour}(a) and (b) where we show the contour plot for integrable point. We observe that the tDMRG with disentangler affords us an improvement in the accessible timescale - the same computational cost (the contour lines) is reached at a later time if the disentangler is applied, as compared to the case without the disentangler. 
However the improvement is most significant at higher temperatures, where the bond dimensions growth is generally much faster and is therefore a relevant consideration. At low temperature where the bond dimension growth is a lesser issue, the disentangler performs slightly worse.
These observations are consistent with Fig. \ref{fig:sitedata_beta_0_16} and Fig. \ref{fig:sitedata_beta_10_0}.

Finally, we discuss Fig. \ref{fig:contour}(c) and (d) where we study a non-integrable point. The plots suggest a minor gain in accessible time scales from ancilla disentangler. However, at non-integrable points, we find that the timescales required for the transients in the current-current correlaters to decay away is much larger compared to the integrable points; the improvements afforded by the disentangler is not sufficient for us to reach these time scales.

\section{Summary and Outlook}
\label{Summary and Outlook}
The constraints imposed by the extensive number of local conservation laws in integrable systems manifest in many different ways - such as failure to thermalize to a Gibbs ensemble, long term memory of initial states which strongly influence equilibrium steady state properties, unusual features in energy spectral correlations, and anomalous signatures in measurable long term response functions such as conductivities. The precise relation between these features has been extensively investigated in the past several decades. This study uses extensive tDMRG calculations to characterize the anomalous zero frequency peak parametrized by the thermal Drude weight at integrable points of the $\mathbb{Z}_3$ chiral clock model.

The thermal Drude weight is nonzero at all finite temperatures and vanishes as a power law $\sim 1/T^2$ at large temperatures and appears to decay exponentially with $1/T$ at low temperatures. At the critical point (gaplesss) the low temperature Drude weight fits a linear behavior with $T$ till the lowest temperature we studied ($T=0.1$). These observations are consistant with what was observed in the gapped and gapless phases of the XXZ model. 

The numerically obtained Drude weights (from current correlators) shows excellent agreement with the Mazur bound based on the simplest conserved charge $Q^{(2)}$ (Eq. \ref{Q2}) derived from the transfer matrix and suggests that the thermal current has finite overlap only with one conserved charge. Explicit calculations using the next conserved charge $Q^{(3)}$ is consistent with this.

The thermal current operator is however not identical to the conserved charge $Q^{(2)}$. 
This corresponds to a scenario where part of the energy current is conserved manifesting as the finite Drude weight. The difference between the current and the conserved charge results in a non-trivial finite frequency part to $\kappa_{\rm reg}$. This is unlike the case of the XXZ model where the current is a conserved charge and $\kappa_{\rm reg}=0$.

The regular part of the dynamical conductivity shows a wide peak at frequencies which could be correlated with the energies of the simplest domain walls. We also find a small finite conductivity near zero frequency suggesting the possibility of a diffusive component to the energy transport in addition to the ballistic part. 

We used ancilla disentangler to increase the maximum times upto which we could compute the correlators. For the integrable points, this allowed us to efficiently access the asymptotically long time behavior of the correlator. However, for the non integrable points we checked, the improvements afforded by the method were insufficient to access long time features of the corelators. More careful choice of the non-integrable parameter regime, use of other disentangler schemes \cite{hauschild2018finding}, alternate computing schemes \cite{unfried2023fast} or truncation methods \cite{white2018quantum,von2022operator} may be helpful in accessing longer time scales.

In this work, we have characterized the anomalous transport properties at the chiral but time reversal invariant points in the integrable parameter space of the model. 
The similarities of our results with the XXZ model highlights the universality of the relation between integrability and anomalous transport, and suggests similar qualitative features for the time reversal broken ($\phi\neq 0$) regime as well. However it will be interesting to understand how broken time reversal invariance manifest in the anomalous transport features. The model also provides an interesting arena to study dynamics and transport at superintegrable\cite{albertini1989commensurate,mccoy1990excitation} points and effects of integrability breaking perturbations on them.

\acknowledgments
We thank R Sensarma and T Sadhu for useful discussions and K Damle and P Fendley for clarifications on previous works. We are grateful to D Dhar for pointing us to Ref~\cite{Anjan1996}.
SGJ thanks Sanjukta Kundu for help with accessing missing figures of Ref \cite{Anjan1996}. SGJ acknowledges Tata Institute for Fundamental Research, Mumbai for their hospitality during the completion of this work.We also thank National Supercomputing Mission (NSM) for providing computing resources of ‘PARAM Brahma’ at IISER Pune, which is implemented by C-DAC and supported by the Ministry of Electronics and Information Technology (MeitY) and
Department of Science and Technology (DST), Government of India.

\appendix
\renewcommand{\thefigure}{A.\arabic{figure}}
\renewcommand{\theHfigure}{A\arabic{figure}}
\setcounter{figure}{0}

\section{Transfer matrix and conserved charges\label{app:TransferMatrix}}
In this section, we give an outline for the derivation of the conserved charges of quantum $\mathbb{Z}_{3}$ chiral clock model from the transfer matrix for a classical 2D chiral clock model. Here we borrow the results for the transfer matrix from Ref \cite{AuYang1987CommutingTM}.

The energy density of the classical 2D model is \cite{ostlund1981incommensurate}
\begin{equation}
	\label{stat mech model}
	\mathcal{E}=-2 {\rm Re}\sum_{j,k}E^v Z_{j,k}Z^{\dagger}_{j+1,k}+E^h Z_{j,k}Z^{\dagger}_{j,k+1}
\end{equation} 
where $E^v$ and $E^h$ are complex coupling strengths along vertical and horizontal directions of the 2D square lattice, respectively and $Z$ takes values from $\{1,\omega, \omega^2\}$. $(j,k)$ label site coordinates on the 2D lattice.

The transfer matrix (with periodic boundary conditions) can be written as
\begin{equation}
	\label{Transfer matrix}
	T_D=\sum_{j=0}^{2}(\sigma^{\dag}_M)^j l_j
 \left[ \prod_{k=-M}^{M-1}(L_k L_{k+\frac{1}{2}})\right]
 \,L_M(\sigma_M)^j 
\end{equation}
where the sites are indexed from $[-M,M]$.
\begin{align}
L_k(u)=&\sum_{n=0}^{2}\bar{l}_n(u) \tau_k^n\nonumber\\
L_{k+\frac{1}{2}}(u)=&\sum_{n=0}^{2} l_n(u)(\sigma_k \sigma_{k+1}^{\dag})^n\nonumber
\end{align}
The weights in the transfer matrix, $\bar{l}_1,{l}_1,\bar{l}_2,{l}_2$ parametrize $E^v, E^{v*}, E^h, E^{h*}$ of the statistical mechanics model and $l_0=\bar{l}_0=1$.

For a certain parameters, $\mathbb{Z}_{3}$ chiral clock model becomes Bethe ansatz integrable. This implies choice of functions $l,\bar{l}$ such that the transfer matrices satisfy Lax pair conditions $[T(u),T(u')]=0$ and $[T(u),H]=0$ where $H$ is a quantum Hamiltonian. 

Ref \cite{AuYang1987CommutingTM} considered functions of the form $l_{1,2}=\alpha_{1,2} u+O(u^2)$ and $\bar{l}_{1,2}=\bar{\alpha}_{1,2} u+O(u^2)$ in which case $H$ matches the quantum chiral clock Hamiltonian (Eq. \ref{Hamiltonian}). The Lax pair conditions can be shown to imply (see Ref \cite{AuYang1987CommutingTM} for details)
\begin{align}
	\label{constraint eqs}
	\alpha_m\sum_{k=0}^2 \frac{\bar{S}_{m+k}}{\bar{S}_{k}} \omega^{-nk}=\bar{\alpha}_n\sum_{k=0}^2 \frac{{S}_{n+k}}{{S}_{k}} \omega^{-mk}
\end{align}
for $m,n \in \{1,2\} $ where
\begin{align}
	\bar{S}_m=\sum_{k=0}^{2} \omega^{mk}\bar{l}_k \quad ,
	{S}_m=\sum_{k=0}^{2} \omega^{mk}{l}_k
\end{align}
which can be obtained from the star-triangle relation. These have solutions provided
\begin{equation}
	\label{integrability condition alpha}
	\frac{\alpha_1^3+\alpha_2^3}{\alpha_1 \alpha_2} = \frac{\bar{\alpha}_1^3+\bar{\alpha}_2^3}{\bar{\alpha}_1 \bar{\alpha}_2}
\end{equation}
These are equivalent to the integrability condition Eq.~\ref{integrability Jf} using the definitions in Eq.~\ref{alpha definition} for the chiral clock model. In the rest of the calculations we restrict to the case where $\bar{\alpha}_1=\bar{\alpha}_2=\bar{\alpha}$ which corresponds to the case where $\phi=0$ as was assumed throught this work.

Logarithm of the transfer matrix is the generator of an infinite set of local conserved charges {\it i.e.}
\begin{align}
	\label{generator of conserved charges}
	Q^{(k)} =\lim_{u \rightarrow 0}\frac{{\rm d}^k}{{\rm d} u^k}\ln T(u)
\end{align}
where $Q^{(k)}$ is the $k^{\rm th}$ conserved charge. 
In the rest of this section, we calculate $Q^{1}$, $Q^{2}$, $Q^{3}$.

At lowest order we get $Q^{(1)}$ to be the quantum Hamiltonian $H$ in Eq. \ref{Hamiltonian}.
To obtain the higher order terms, we extend the expansion for the functions $l,\bar{l}$ as follows
\begin{align}
	\label{llbardefinition}
	l_n(u)&=\alpha_n u +\beta_n u^2 +\gamma_n u^3+ O(u^4)\nonumber\\
	\bar{l}_n(u)&=\bar{\alpha}_n u +\bar{\beta}_n u^2 +\bar{\gamma}_n u^3+ O(u^4)
\end{align}
for $n=1,2$. Note that $\beta$,$\gamma$ are not complex conjugates of $\bar{\beta}$,$\bar{\gamma}$. We solve for these additional parameters by inserting Eq. \ref{llbardefinition} in Eq. \ref{constraint eqs} and equating coefficients of powers of $u$ on both sides. 

Equating coefficients of the lowest order in $u$ in Eq. \ref{constraint eqs}, we get the following solution parametrized by one free parameter:
\begin{gather}
	\bar{\beta}_1=\omega (\frac{\bar{\alpha}}{\alpha_2}\alpha_1^2-\bar{\alpha}^2)+(\frac{\bar{\alpha}}{\alpha_2})\beta_2\nonumber\\
	\bar{\beta}_2=\omega^2 (\frac{\bar{\alpha}}{\alpha_2}\alpha_1^2-\bar{\alpha}^2)+(\frac{\bar{\alpha}}{\alpha_2})\beta_2\nonumber\\
	2(\alpha_1 \beta_2 -\alpha_2 \beta_1)=\alpha_1 ^3-\alpha_2^3.\label{beta condition}
\end{gather}
We can plug in solutions of Eq. \ref{beta condition} in the Eqs. \ref{llbardefinition}, subsequently expand Eq. \ref{Transfer matrix} and use Eq. \ref{generator of conserved charges} to get the the conserved charge $Q^{(2)}$ explicitly (this does not depend on the undetermined parameters $\gamma$s). Demanding Hermiticity fixes the free parameter in Eq. \ref{beta condition} and we obtain,
\begin{align}
	Q^{(2)}=\sum_k [& I_k+\imath\left(\frac{1}{2}+\omega \right)
	(\frac{ \alpha_1^2 \bar{\alpha}}{\alpha_2}-\bar{\alpha}^2)\tau_k+ \nonumber \\
	& \imath \left(\frac{1}{2}+\omega^2 \right)
	(\frac{ \alpha_1^2 \bar{\alpha}}{\alpha_2}-\bar{\alpha}^2)\tau^{\dag}_k
	]
\end{align}

Similarly we can constrain $\gamma$ by equating the subsequent order in $u$ in Eq. \ref{constraint eqs} resulting in the following conditions involving $\gamma$ and $\bar{\gamma}$
\begin{gather}
		(\omega-1)(\bar{\alpha}\gamma_2-\alpha_2 \bar{\gamma_2}) =2(2+\omega)\alpha_2\bar{\alpha}\bar{\beta_1}+(\omega-4)\alpha_2 \bar{\alpha}^3 \nonumber \\
		-\alpha_1 \bar{\alpha}(\alpha_2^2(\omega-4)+2\beta_1(2+\omega)) \nonumber \\
            (\omega+2)(\bar{\alpha}\gamma_2-\alpha_2 \bar{\gamma_1}) =2(\omega-1)\alpha_2\bar{\alpha}\bar{\beta_2}+(\omega+5)\alpha_2 \bar{\alpha}^3 \nonumber \\
		-\alpha_1 \bar{\alpha}(\alpha_2^2(\omega+5)+2\beta_1(\omega-1))\nonumber \\
            3 (\alpha_2 \gamma_1-\alpha_1 \gamma_2)=2\alpha_2^2 \beta_2 (-3 \omega )-3(1+2\omega) \alpha_1 \alpha_2 (\bar{\alpha}^2-2 \bar{\beta_1})\nonumber \\
		-\alpha_1^2(2 \beta_1 (-3\omega^2)-3 \alpha_2^2(1+2\omega))
\end{gather}
These have a one parameter set of solutions (for $\gamma$s; the other parameters were determined in the previous step) which we fix by demanding Hermiticity of $Q^{(3)}$(its explicit form does not depend on higher order terms omitted in Eq. \ref{llbardefinition}) which has the form:
\begin{align}
	\label{Q3}
	Q^{(3)}=\sum_k [&A \sigma_k (\tau^{\dag}_k+\tau^{\dag}_{k+1}) \sigma^{\dag}_{k+1}+
	B \sigma_k (\tau_k+\tau_{k+1}) \sigma^{\dag}_{k+1}-\nonumber \\
	&C (\tau_k\sigma_k  \tau_{k+1}+\tau^{\dag}_k \sigma_k  \tau^{\dag}_{k+1})  \sigma^{\dag}_{k+1}+\nonumber\\
	&C \sigma_k (\tau^{\dag}_k \tau_{k+1}+\tau_k \tau^{\dag}_{k+1} ) \sigma^{\dag}_{k+1}+\nonumber\\
	&D \sigma_k (\tau_{k+1}+\tau_{k+1}^{\dag}) \sigma^{\dag}_{k+2}+\nonumber\\
	&E \sigma_k  (\sigma_{k+1} \tau_{k+1} +\tau^{\dag}_{k+1} \sigma_{k+1} )\sigma_{k+2}+\nonumber\\
	&F \tau_k+G \sigma_k\sigma^{\dag}_{k+1} \nonumber \nonumber\\+
	h.c
	]
\end{align}
where 
\begin{equation*}\label{eq1}
	\begin{gathered}
	A=\frac{\omega^2}{2}\bar{\alpha}\alpha_2^2-\omega^2\bar{\alpha}^2\alpha_1,\\
	B= \frac{\omega}{2}\bar{\alpha}\alpha_2^2-\omega\bar{\alpha}^2\alpha_1,\\
	C=\bar{\alpha}^2 \alpha_1,\\
	D=-\bar{\alpha} \alpha_1^2,\\
	E=\omega \bar{\alpha} \alpha_1 \alpha_2,\\
	F=3 \bar{\alpha} \alpha_1 \alpha_2,\\
	G=\frac{3}{2} \bar{\alpha} \alpha_2^2 +\bar{\alpha}^2 \alpha_1 + \frac{1}{2}\alpha_1^2\alpha_2.
	\end{gathered}
\end{equation*}
We find that both these charges $Q^{(2)}$ and $Q^{(3)}$ have terms with support on up to $3$ sites. These are orthogonal under the norm $\langle,\rangle_{\rho}$ defined in the main text at all temperatures we tested.

\section{Convergence of real-time data from tDMRG \label{app:convergence}}
\begin{figure}[H]
	\includegraphics[width=\columnwidth]{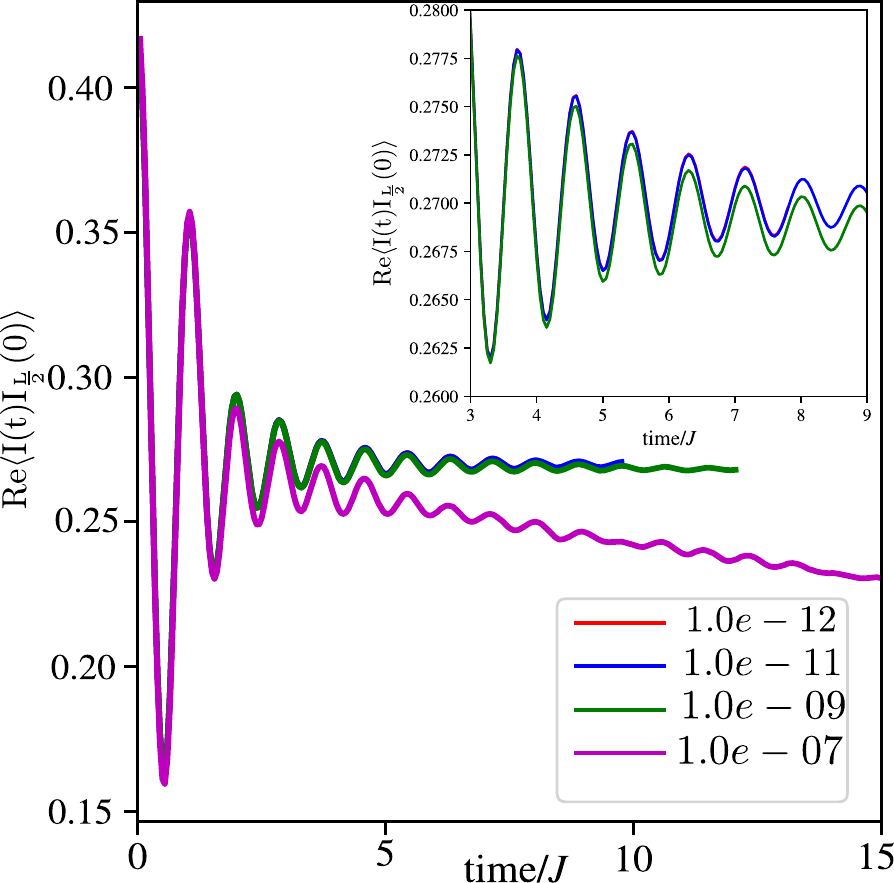}
	\caption{Convergence of $\mathrm{R e}\langle I(t)I_{L/2}(0) \rangle$ with decreasing truncation cutoff. $J=1$, $f=\frac{2}{5}$, $\theta=\cos^{-1}(\frac{2}{5})$, $T=1$ . Maximum bond dimension is 900 in all cases. (Inset) Magnified version of the $\mathrm{R e}\langle I(t)I_{L/2}(0) \rangle$ data. Notice the blue and red curves overlap.} 
	\label{fig:cutoff_convergence}
\end{figure}

Truncation cutoff ($\epsilon$) affects the computational cost and accuracy of tDMRG calculation. We verified the convergence of our current-current correlator by comparing results at different $\epsilon$. For $T=1$, we observe in Fig.~\ref{fig:cutoff_convergence} that the truncation cutoff $10^{-9}$ is sufficient. At larger cutoff (e.g $1e-7$), the $\rm{Re} \langle I(t)I_{\frac{L}{2}} \rangle$ decreases fast from around time 5 indicating reliability of the result only upto small timescales. On the other hand, for $\epsilon \leq 10^{-9}$ the asymptotic behaviour shows very small difference at long timescale indicating convergence of result.

\begin{figure}[H]
	\includegraphics[width=\columnwidth]{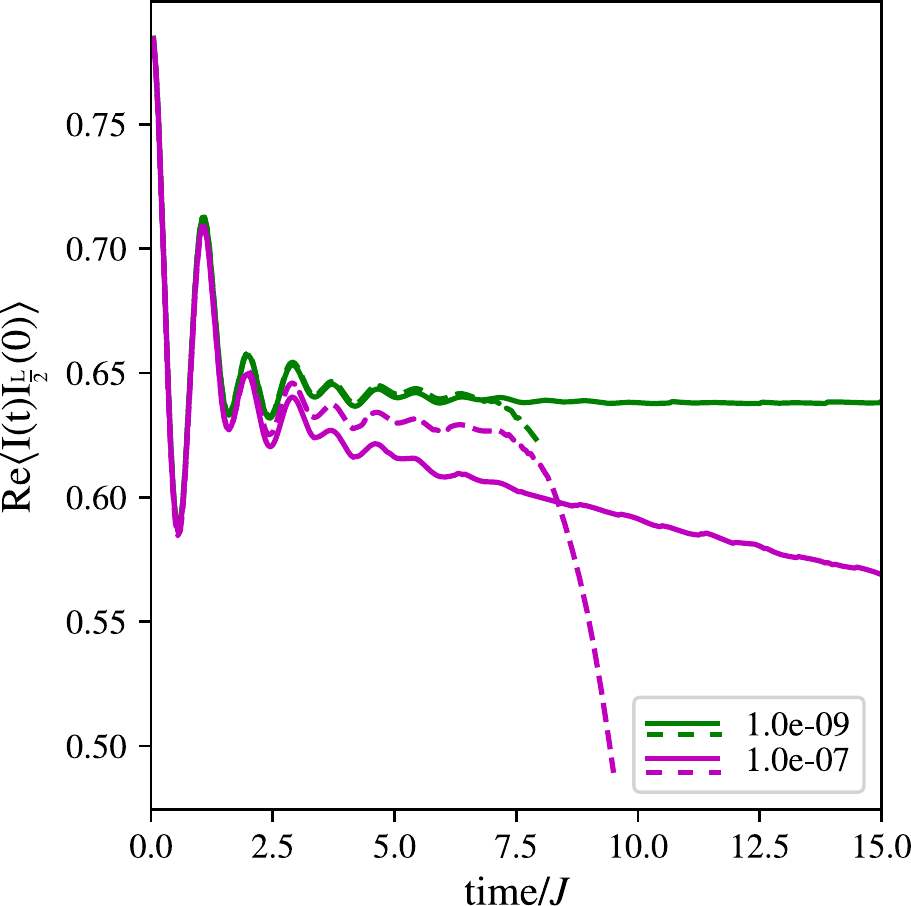}
	\caption{Similar to Fig. \ref{fig:cutoff_convergence} but shows the effect of ancilla disentangler. Solid line is with tDMRG with disentangler unitary and dotted line is without application of ancilla disentangler. Maximum bond dimension is 900 in all cases. $J=1$, $f=\frac{2}{5}$, $\theta=\cos^{-1}(\frac{2}{5})$, $T=2$ . }
	\label{fig:ancilla_convergence}
\end{figure}

In Fig. \ref{fig:ancilla_convergence}, we observe that for the same truncation cutoff and bond dimensions, ancilla disentangler produces reliable results for significantly longer timescales.

\section{Choice of finite temperature disentangler unitary\label{app:disentangler}}

As noted in the main text, invariance of the reduced density matrix under unitary transformations of the $3^L$ dimensional ancilla Hilbert space allow us to choose an optimal `disentangler' unitary transformation of the ancilla that minimize entanglement growth. A search within the $3^{2L}$ dimensional parameter space of the unitary group $U(3^L)$ is computationally impossible, so we consider a one parameter set of anicilla unitaries of the form $e^{\imath h(\theta_{\rm anc})}$ where $h(\theta_{\rm 
 anc})$ has the same form as the clock model Hamiltonian but with a parameter $\theta_{\rm{anc}}$ replacing $\theta$ of the physical Hamiltonian.
\begin{figure}[H]
        \label{fig:theta_scan_merged}
	\includegraphics[width=\columnwidth]{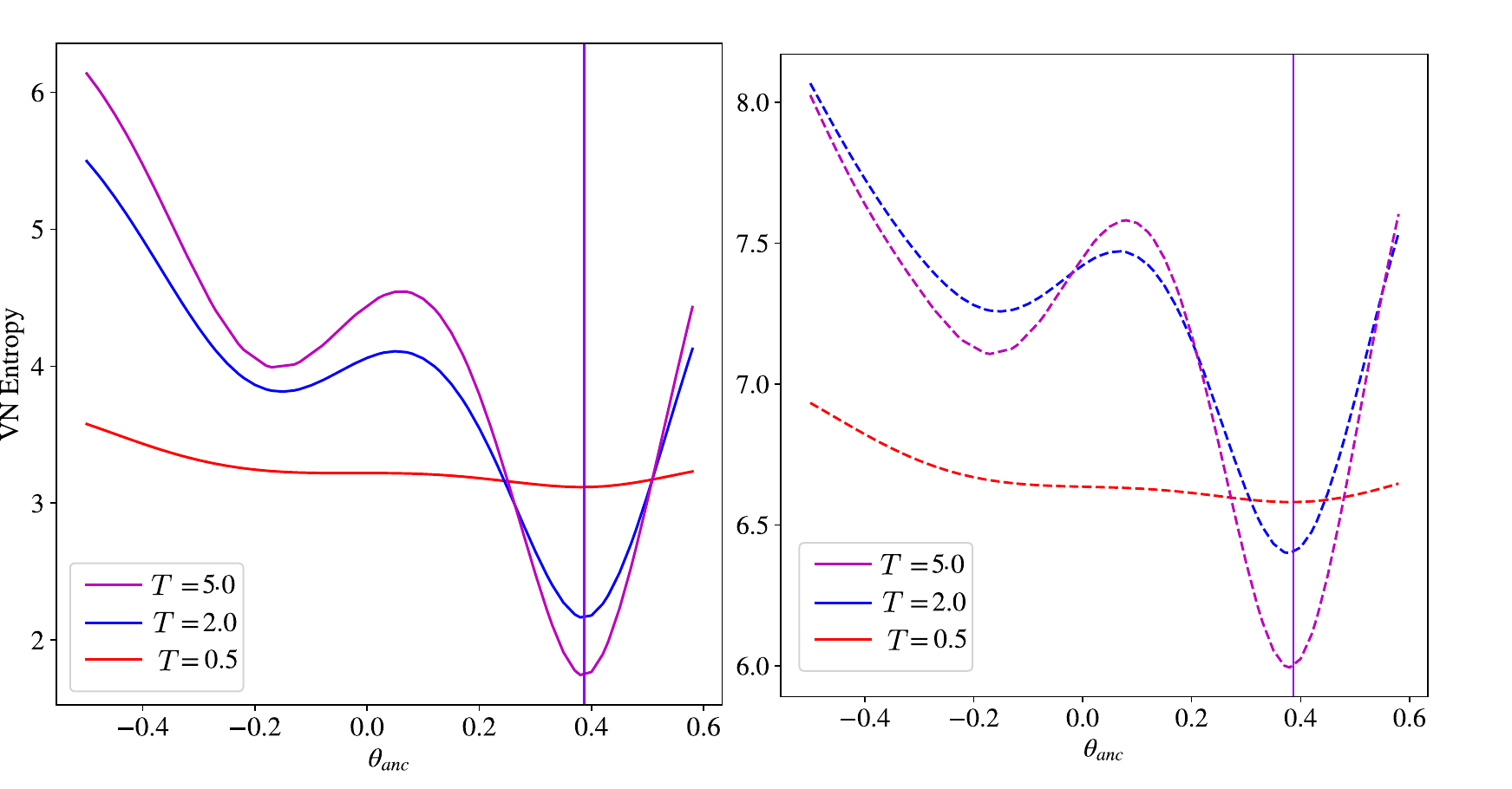}
	\caption{Optimal value for $\theta_{\rm  anc}$ decided from the bipartite Von-Neumann entropy on bonds near the quench site. The violet vertical line corresponds to $\theta_{\rm anc} = \theta $. The solid lines correspond to entanglement growth in thermal state and the dotted lines correspond to entanglement growth in thermal state evolved after quench. The entanglement is measured after time 2.5. Calculations are done at integrable point corresponding to $f=2/5$ on a system of 30 physical sites. 
    }
\end{figure}
We observe that for $\theta_{\rm  anc}=\theta$ (Fig. \ref{fig:theta_scan_merged}), the entanglement growth (as measured by the bipartite Von Neumann entropy) on the bonds near the quench site is minimum at any temperature. The disentangler is most effective at high temperature. Similar conclusion holds for non-integrable points too.

\section{Consistency check of conductivity data\label{app:consistency}}
\begin{figure}[H]
 	\includegraphics[clip,width=\columnwidth]{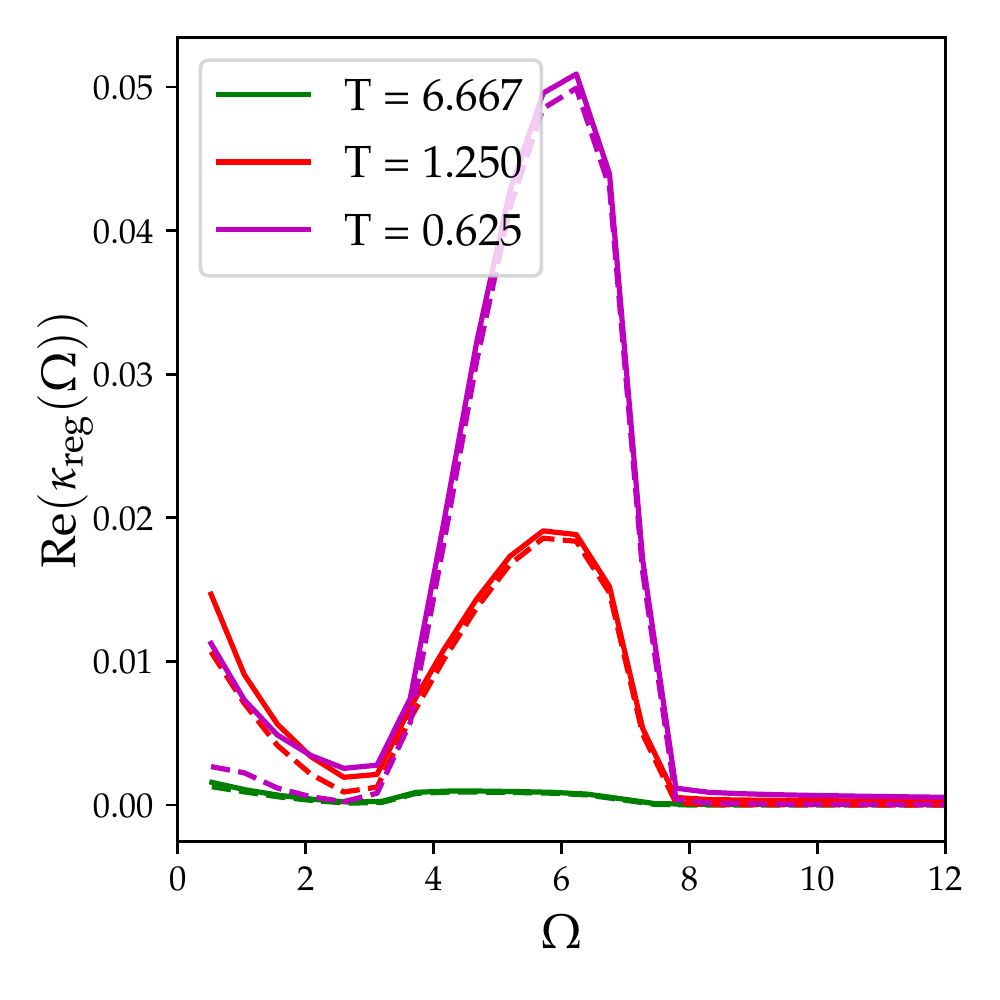}%
 	\caption{$\mathrm{Re}\ \kappa_{\rm reg}$ obtained from  eq.\ref{conductivity current current correlation} is shown in solid line and the same from \ref{reg imaginary conductivity current current correlation} is shown in dotted line. Notice both curves overlap for large $\Omega$ while deviation can be observed for small $\Omega$.   
}
 	\label{fig:real_im_sigma}
 \end{figure}

As discussed in section \ref{drude weight}, we can evaluate the thermal conductivity using two equivalent expressions ({\it i.e.} Eq. \ref{conductivity current current correlation} and Eq. \ref{reg imaginary conductivity current current correlation}). They should result in same values if we have data for correlator at all time. However, tDMRG calculation can provide reliable data only up to a finite timescale. Consequently, we incur a ``finite-time error" in the discrete Fourier transform required for thermal conductivity calculation ( Eq. \ref{reg imaginary conductivity current current correlation}). We can estimate this error from the difference in the values obtained from Eq. \ref{conductivity current current correlation} and Eq. \ref{reg imaginary conductivity current current correlation}. In Figure \ref{fig:real_im_sigma}, we provide the data for 3 different temperature at a single point ($f=2/5$) on integrable line. It can be readily observed that the estimates from Eq. \ref{conductivity current current correlation} (solid line) and Eq. \ref{reg imaginary conductivity current current correlation} (dotted line) match quite well for most finite frequency region with significant difference between them only at low frequency. This is expected as low-frequency regime gets most affected by long time data. However, the estimated errror obtained here is order of magnitude smaller than Drude weight.

\typeout{} 
\bibliographystyle{unsrt}
\bibliography{refs.bib}	
\end{document}